\documentclass[onecolumn,showpacs,amsmath,amssymb,preprint,tightenlines,10pt]{revtex4-1}
\pdfoutput=1
\usepackage{graphicx,anysize}
\usepackage{amsmath,mathrsfs,amsbsy}
\usepackage{amsmath}
\usepackage{amssymb}
\usepackage{psfrag}
\usepackage{threeparttable}
\usepackage{float}
\usepackage{color}
\usepackage{varioref}
\usepackage{epic}
\usepackage{eepic}
\usepackage{makeidx}
\usepackage{epsfig}
\usepackage{supertabular}
\usepackage{amsmath,setspace}
\usepackage[english]{babel}

\newcommand{\radt}[2][]{${\cal R}_{\rm tether}$#2}
\newcommand{\lent}[2][]{${\cal L}_{\rm tether}$#2}
\newcommand{\fort}[2][]{${\cal F}_{\rm tether}$#2}

\begin{document}

\title{Thermodynamic Free Energy Methods to Investigate Shape Transitions In Bilayer Membranes}

\author{N. Ramakrishnan} 
\email{ramn@seas.upenn.edu}
\affiliation{Department of Bioengineering, 
	University of Pennsylvania, 
	Philadelphia, PA, 19104}
\email{ramn@seas.upenn.edu}
\author{Richard W. Tourdot}
\affiliation{Department of Chemical and Biomolecular Engineering, 
	University of Pennsylvania, Philadelphia, PA, 19104 }
\email{tourdotr@seas.upenn.edu}
\author{ Ravi Radhakrishnan}
\affiliation{
	Department of Bioengineering, 
	Department of Chemical and Biomolecular Engineering, \\
	Department of Biochemistry Biophysics,
	University of Pennsylvania, Philadelphia, PA, 19104 }
\email{rradhak@seas.upenn.edu}

\date{\today}
\begin{abstract}
The conformational free energy landscape of a system is a fundamental thermodynamic quantity of importance particularly 
in the study of soft matter and biological systems, in which the entropic contributions play a dominant role. While  
computational methods to delineate the free energy landscape are routinely used to analyze the relative stability of 
conformational states, to determine phase boundaries, and to compute ligand-receptor binding energies its 
use in problems involving the cell membrane is limited. Here, we present an overview of four different free energy 
methods to study morphological transitions in bilayer membranes, induced either by the action of curvature remodeling 
proteins or due to the application of external forces. Using a triangulated surface as a model for the cell membrane 
and using the framework of dynamical triangulation Monte Carlo, we have focused on the methods of Widom insertion, 
thermodynamic integration, Bennett acceptance scheme, and umbrella sampling and weighted histogram analysis.  We have
demonstrated how these methods can be employed in a variety of problems involving the cell membrane. Specifically, we 
have shown that the chemical potential, computed using Widom insertion, and the relative free energies, computed 
using thermodynamic integration and Bennett acceptance method, are excellent measures to study the transition from 
curvature sensing to curvature inducing behavior of membrane associated proteins. The umbrella sampling and WHAM 
analysis has been used to study the thermodynamics of tether formation in cell membranes  and the quantitative 
predictions of the computational model are in excellent agreement with experimental measurements. Furthermore, we also 
present a method based on WHAM and thermodynamic integration to handle problems related to end-point-catastrophe that 
are common in most free energy methods.
\end{abstract}

\preprint{To appear in Int. J. Adv. Eng. Sci. \& Appl. Math.}

\maketitle

\section{Introduction}
Surfactant molecules self assemble into mesoscale structures (characteristic lengths are in the order of hundreds of nanometer) 
when their concentration in an aqueous solvent exceeds a threshold value, generally called the critical micelle concentration 
(CMC). Examples of these mesoscale entities include simple structures like a monolayer of surfactants at the air-water/air-oil 
interface or more complex structures like a micelle and a bilayer of surfactants in the bulk. The stability of a given mesoscale 
structure is in turn is governed by the geometry and chemistry of the individual surfactant molecules ~\cite{Israelachvili:2011wy}. 
Characteristic energies of a self assembled surfactant interface are comparable to the thermal energy $k_{B}T$, where $k_{B}$ is 
the Boltzmann constant and $T$ is the equilibrium temperature, and a result the spatial organization of the molecules, which is 
characterized at the mesoscale by the morphology and topology of the interface, is susceptible to thermal fluctuations in the 
solvent. 

A similar but a more complex system that is of importance to cell biology is the lipid bilayer membrane, formed by the 
self assembly of lipid molecules, which defines the outer boundaries of most mammalian cells and their organelles. 
Lipid molecules are fatty acids synthesized within the cell and like a surfactant molecule they also have a hydrophilic 
head group and a hydrophobic tail \textemdash commonly occurring lipids include glycerol based lipids such as DOPC, DOPS 
and DOPE, sterol based lipids like cholesterol, and ceramide based lipids like sphingomyelin~\cite{Escriba:2008hb}. The 
cell membrane is formed by the self assembly of these different types of lipid molecules along with other constituents 
namely proteins and carbohydrates, and the composition of these building blocks differ across different cell 
membranes~\cite{Singer:1972uq,Edidin:2003fv,Engelman:2005bo}. Being the interface of the cell, the lipid membrane 
plays a dominant role in a number of biophysical processes either by virtue of its surface chemistry at the molecular scale or 
through modulations in its physical properties at the mesoscale: the most obvious examples of the latter include inter- and 
intra-cellular trafficking~\cite{Conner:2003gm,Doherty:2009fz,Ewers:2011jj,Canton:2012dl}, membrane mediated aggregation of cell 
signaling molecules~\cite{Kholodenko:2006cr,Sorkin:2009cb} and cell 
motility~\cite{Sheetz:2001ed,Ananthakrishnan:2007ux,Keren:2011hg}. Hence, it is natural to expect an inherent feedback 
between the physical properties of the cell membrane and the biophysical processes it mediates. The primary aim of this 
article is to review theoretical and computational approaches at the mesoscale that can be used to develop an understanding of this 
feedback. In particular, our focus is to show how thermodynamic free energy methods employed in a variety in a contexts 
in condensed matter physics can be applied to the theoretical models for membranes at the mesoscale.

In equilibrium statistical mechanics, the ground state of a system whose intensive or extensive variables are coupled to the 
environment, and hence can exchange for instance heat or area or volume or number with the bath, is governed by its thermodynamic 
potential which is also called  the free energy of the system. The various thermodynamic observables can be determined by measuring 
the suitable thermodynamic potential that depends on the ensemble in which the system is defined~\cite{Chaikin:2000td}. Excellent introduction to the implementations and applications of the various free energy methods for molecular systems is provided by Frenkel and Smit~\cite{Frenkel:2001}.

\section{Continuum models for cell membranes}
The spatial and temporal resolution of the various biophysical processes observed in cell membranes can be classified into two 
broad classes, namely (a) biochemical processes in which the dynamics of the system is primarily determined by the chemistry of the 
constituent molecules and (b) biophysical processes where collective phenomena and macroscopic physics govern the behavior of the 
membrane. These two class of processes have disparate time and length scales. The large separation in the time and 
length scales allows one to decouple the slower degrees of freedom from the faster ones and this feature can be 
exploited in constructing physical models at multiple scales for the cell membrane. Molecular scale models such as 
all-atom or coarse grained molecular dynamics are faithful to the underlying chemistry and are hence more appropriate 
for investigating membrane processes in the sub cellular length and nanoscopic time scales. In the other 
limit, phenomenology based field theoretic models neglect the membrane dynamics at the nanoscale and instead focus on how the 
collective effects of these molecular motions manifest at length and time scales comparable to those accessed in conventional 
experiments like light microscopy and mechanical measurements of cells. More rigorous discussions on the formulation of 
multiscale models for membranes can found in a number of review articles on this 
topics~\cite{Seifert:1997wq,Tieleman:1997ve,Venturoli:2006kk,Ayton:2010gv,Shinoda:2012ks,Bradley:2013gb,Ramakrishnan:2014ed,Deserno:2014cv}. In this article, we will use the thermodynamic formalism of membrane biophysics to demonstrate how free energy 
methods can extended to the study of diverse class of problems involving the cell membrane at the mesoscale.

The phenomenology based approach focuses primarily on the conformational states of the bilayer membrane at length scales ($>100$  
nm) that are large compared to the thickness of the membrane ($\sim$ 5 nm). In this approach  the membrane is treated 
as a thin elastic sheet of a highly viscous fluid with nearly constant surface area (the number of lipids under 
consideration is  assumed to be constant). This sheet is representative of the neutral surface of a membrane bilayer: it 
is defined as the cross sectional surface in which the in-plane strains are zero upon a bending transformation, see 
references ~\cite{Seifert:1997wq,Ramakrishnan:2014ed,Deserno:2014cv} for details. The thermodynamic weights of the 
conformational state of the membrane is governed by the well known Canham-Helfrich 
energy functional~\cite{Canham:1970p61,Helfrich:1973td} commonly written as,
\begin{equation}
\mathscr{H}= \underset{{\bf S}} \int d{\bf S} \left\{\frac{\kappa}{2} (2H-H_{0})^{2} +\kappa_{G} G +\sigma \right\}+ \underset{V} 
\int dV \Delta p.
\label{eqn:can-Helf-gen}
\end{equation}
If $c_{1}$ and $c_{2}$ are the principal curvatures at every point on the membrane surface ${\bf S}$ then  $H=(c_{1}+c_{2})/2$ and 
$G=c_{1}c_{2}$ are its mean and Gaussian curvatures respectively. The elastic moduli $\kappa$ and $\kappa_{G}$ are the isotropic 
and deviatoric  bending moduli. Experimental measurements on lipid and cell membranes  have estimated their bending stiffness 
$\kappa$ to be in the range $10-100k_{B}T$, with the lower values corresponding to  model membrane structures like giant 
uni-lamellar vesicles. The deviatoric modulus is normally taken to $\kappa_{G}=-\kappa$ but the Gaussian energy term can be 
neglected, by virtue of the Gauss-Bonnet theorem~\cite{doCarmo:1976} if the topology of the membrane does not change during the 
analysis.  The surface area of the membrane $A$ and the  volume $V$ are coupled to their respective conjugate variables namely the 
surface tension $\sigma$ and osmotic pressure $\Delta p$. Reported values of membrane surface tension (combined contributions from 
both the lipids and the underlying cytoskeleton) varies between 3--300 $\mu$N/m depending on the cell 
type~\cite{Shi:2015ho,DizMunoz:2013bi}. The osmotic pressure difference $\Delta p$ is a function of the difference in the osmolyte  
concentration between the inside and outside of the cell.

The spontaneous curvature $H_{0}$ denotes an induced curvature which can arise in a number of context such as defects in lipid 
packing, the presence of intrinsic degrees of freedom in the constituent lipids, interactions of the membrane with non-lipid 
molecules like proteins or nanoparticles and also due to the coupling of the membrane with the underlying cytoskeleton. Since the 
spontaneous curvature is an important parameter in most of our discussions later, it is important to  have a closer look at how its 
impacts the conformational states of the membrane. When $H_{0}=0$, the energy given by the first term in 
eqn.~\eqref{eqn:can-Helf-gen} is quadratic in the mean curvature $H$ and hence the probability of finding a membrane conformation 
with a given curvature $H^{*}$ is a Gaussian peaked around $H=0$, with its width being proportional to the bending stiffness 
$\kappa$~\cite{Nelson:2004vm}. On the other hand, when the membrane has a non-zero spontaneous curvature the peak of the 
probability distribution now shifts to a value $H=H_{0}$ and as a result highly curved membrane regions are observed with much 
larger probabilities. 

For purposes of computer simulations, a number of discretizations based on eqn.~\eqref{eqn:can-Helf-gen} have been introduced in 
the literature. The free energy methods for membranes presented in the later sections are based on the Dynamical Triangulation 
Monte Carlo technique which has been reviewed in brief below.

The two dimensional membrane surface is discretized into an interconnected set of $T$ triangles that  intersect at $N$ vertices 
({\it nodes}) forming $L$ independent links.  The values of $N$, $T$, and $L$ define the topology of the membrane surface in terms 
of Euler characteristic as $\chi=N+T-L$. The degrees of freedom of the discretized membrane are the position vectors of the $N$ 
vertices given by $\{\vec{X}\}=[\vec{x}_{1} \cdots \vec{x}_{N}]$ and the triangulation map given by 
$\{\mathscr{T}\}=[\mathscr{T}_{1}\cdots \mathscr{T}_{T}]$. The discrete form of the elastic Hamiltonian is thus a sum over the 
curvature energies at every vertex in the triangulated surface given by,
\begin{equation}
\mathscr{H}= \sum_{v=1}^{N} A_{v} \left\{\dfrac{\kappa}{2} (c_{1,v}+c_{2,v}-H_{0,v})^{2} +\sigma \right\} + \Delta p V.
\label{eqn:can-Helf-discrete}
\end{equation}
 The index $v$ denotes a vertex on the triangulated surface and $c_{1,v}$ and $c_{2,v}$ are respectively its principal curvatures, 
$H_{0,v}$ is the local spontaneous curvature, and $A_{v}$ denotes the surface area associated with the vertex. The principal 
curvatures are computed using the methods introduced by Ramakrishnan et. al.~\cite{Ramakrishnan:2010hk}. The spontaneous curvature 
at a vertex is  expressed using the general form:
\begin{equation}
H_{0,v}=\sum_{v=1}^{N} C_{0}{\cal D}(v,v^{'}),
\label{eqn:czeroform}
\end{equation}
with $C_{0}$ being the magnitude of the induced curvature and ${\cal D}(v,v^{'})$  the functional form  of the curvature 
contribution at vertex $v^{'}$ due to a curvature field at vertex $v$. The various forms of ${\cal D}(v,v^{'})$ 
relevant in 
different contexts have been discussed in 
references~\cite{Agrawal:2010iu,Ramanan:2011ds,Liu:2012es,Ramakrishnan:2014ed,Tourdot:2014wh}. In this article, we limit our 
discussions  on curvature induced membrane remodeling to protein that have  isotropic curvature fields with a Gaussian profile:
\begin{equation}
{\cal D}(v,v^{'})= \exp\left(- \dfrac{|{\bf x}_{v}-{\bf x}_{v^{'}}|^{2}}{2\epsilon^{2}} \right ).
\label{eqn:curvprofile} 
\end{equation} 
Here $\epsilon^{2}$ denotes the range of a curvature field, i.e. a curvature field defined at a vertex $v$ can induce a 
non-zero spontaneous curvature at a far vertex $v^{'}$. We denote the set of all protein fields as $\{\phi 
\}=[\phi_{1}\cdots \phi_{N}]$.

In addition to the elastic potential given in eqn.~\eqref{eqn:can-Helf-discrete} the membrane vertices are also subjected to a 
repulsive hard-sphere potential in order to enforce self avoidance. If the vertices of the membrane are taken to spheres of 
diameter $a_{0}$ then the length ${\cal E}$ of the links, connecting any two vertices, obey the constraint $a_{0} \le |{\cal E}|< 
\sqrt{3}a_{0}$. Detailed discussions on this topic can be found in reference [cite Elsevier chapter]. The conformational state of 
the triangulated membrane  is given by $\boldsymbol{\eta}=[ \{\vec{X}\},\{\mathscr{T}\},\{\boldsymbol{\phi} \}]$ and the various 
state are sampled using a set of three Monte Carlo moves: (i) a {\it vertex move} in which a randomly chosen vertex is displaced to 
new location that leads to change in state $[ \{\vec{X}\},\{\mathscr{T}\},\{\boldsymbol{\phi} \}] \rightarrow [ 
\{\vec{X}^{'}\},\{\mathscr{T}\},\{\boldsymbol{\phi} \}]$, (ii) a {\it link flip} wherein two previously unconnected nodes of a 
randomly chosen quadrilateral on the triangulated surface are connected to form a new set of triangulation leading to  $[ 
\{\vec{X}\},\{\mathscr{T}\},\{\boldsymbol{\phi} \}] \rightarrow [ \{\vec{X}\},\{\mathscr{T}^{'}\},\{\boldsymbol{\phi} \}]$, and 
(iii) a {\it field exchange} move to simulate diffusion of the protein field in which the protein field at vertex $v$ is exchanged 
with that at vertex $v^{'}$ that leads to a change in state $[ \{\vec{X}\},\{\mathscr{T}\},\{\boldsymbol{\phi} \}] \rightarrow [ 
\{\vec{X}\},\{\mathscr{T}\},\{\boldsymbol{\phi}^{'}\}]$  . The various Monte Carlo moves are accepted using Metropolis 
scheme~\cite{Metropolis:1953in} given by $P_{\rm acc}=\min \left \{ 1, \exp\left (-\beta \left 
(\mathscr{H}(\boldsymbol{\eta}^{'})-\mathscr{H}(\boldsymbol{\eta}) \right )\right )\right\}$, where $\beta=1/k_{B}T$ denotes the 
temperature of the system. More details on the implementation and usage of Dynamical Triangulation Monte Carlo techniques can be 
found in reference~\cite{Ramakrishnan:2014ed}.

\section{An overview of free energy methods to study membrane deformations} \label{sec:overview}
As noted in the introduction, the thermodynamic free energy of a system is a fundamental quantity in equilibrium statistical 
mechanics since it contains all the information  about the thermodynamic variables of the system. However, the free energy 
landscape of many body systems is a very complex quantity and the complexity arises primarily from the large number of degrees of 
freedom associated with such systems \textemdash~ for example an $N$ particle system in one dimensions has a conformational free 
energy landscape that is $N$ dimensional. In most problems in  condensed matter physics, computational biology, and computational 
chemistry the free energy landscape in the conformational space of the atoms/molecules is an over representation of the system and 
hence the problem of large number of dimensions can be overcome through coarse graining or representing the system in terms of a 
fewer macroscopic variables generally called order parameters. The use of the computational methods to delineate free energy 
landscape is highly optimal in such coarser representations. The problem of protein induced curvature remodeling of membranes is 
one such problem that is amenable to the use to free energy methods. The partition function of the triangulated membrane 
surface with $n$ protein fields is defined in terms of the thermodynamic state 
$\eta$ as,
\begin{equation}
Q_{n}=\sum_{\phi \in \{\boldsymbol{\phi}\}} \sum_{\mathscr{T} \in \{\mathscr{T}\}} \int d \{{\bf X}\} \,\exp 
\left(-\beta 
\mathscr{H}({\bf x},\mathscr{T},\phi) \right).
\label{eqn:partfunc}
\end{equation}
The partition trace in eqn.~\eqref{eqn:partfunc} is performed over all vertex positions and triangulations of the discrete surface 
and also over all possible configurations of the $n$ proteins on the membrane. The absolute free energy of the membrane with $n$ 
proteins is thus:
\begin{equation}
{\cal F}_{n}=-k_{B}T \ln Q_{n}.
\label{eqn:freeen-def}
\end{equation}

Computing the absolute free energy $F_{n}$ requires the calculation of the partition function $Q_{n}$ which is a problem that 
requires extensive sampling of the infinitely large conformational states of the membrane-protein system \textemdash~ Metropolis 
Monte Carlo is not suitable for such purposes since it can only sample the configurational states close to the free energy minimum 
and the higher energy states can instead be sampled using rare event Monte Carlo techniques which are described later in the 
context of umbrella sampling. On the other hand, the relative free energy of state $m$ with respect to state $n$, i.e. the quantity 
${\cal F}_{m}-{\cal F}_{n}$, does not require the knowledge of $Q_{n}$ and can be computed with less expensive computations.

In the remainder of this section, we will describe three different methods namely the Widom insertion, thermodynamic integration 
(TI), and Bennett Acceptance method (BAM) to essentially perform the same calculation \textemdash~ to compute  $\Delta {\cal 
F}={\cal F}_{n+1}-{\cal F}_{n}$, the free energy difference between a membrane with $n$ and $n+1$ proteins.
\subsection{Widom insertion technique}
 The Widom particle or test-particle insertion method is a computational technique to probe the chemical potential of a system 
\cite{Widom:1963fl}. It is known from statistical mechanics that the total chemical potential of a system is defined as 
the change in its free energy in response to a change in the system size. In the case of membrane-protein systems, the 
total chemical potential of the membrane with $n$ proteins ($\mu_{P}$) is essentially the required  free energy 
difference, i.e. $\mu_{P} = \Delta {\cal F}/{\Delta n}$. 
 
 The total chemical potential can be separated into an ideal and an excess part such that  \begin{equation}
\mu_{P}=\mu_{P}^{id}+\mu_{P}^{ex}.
\end{equation}
   
 The excess part of the chemical potential $\mu_{P}^{ex}$ can be computed using Widom insertion technique in which a virtual 
test(ghost) protein field is inserted at a randomly chosen location on the membrane. The configurational component of the ideal 
part can be shown to be  $\mu_{P}^{id}=k_{B}T \ln \rho$, where $\rho$ is the protein density. If $\Delta {\mathscr{H}}$ 
be  the change in the elastic energy due to the insertion of a test 
protein field then the excess chemical potential is given by,
\begin{equation}
\mu_{P}^{ex} = -k_B T \ln{\int{ \left \langle \exp(-\beta \Delta {\mathscr{H}}) \right \rangle P_{\rm uniform} 
(s_{n+1}) 
ds_{n+1}}}.
\label{eqn:muex}
\end{equation}
The ensemble average $\langle  \cdot \rangle$ is taken over the configurational space of the partition function, see 
eqn.~\eqref{eqn:partfunc}. Here, $s_{n+1}={\bf x}_{p}$, with $p=n+1$, is the position of the $n+1^{\rm th}$ protein field on the 
membrane surface, and $P_{\rm uniform} (s_{n+1})$ denotes a uniform probability distribution from which the coordinates of the 
$n+1^{\rm th}$ particle/field is drawn. The integral over $s_{n+1}$ amounts to the sum over all Widom test particle/field insertion 
trials, and $P_{\rm uniform} (s_{n+1})$ equals the reciprocal of the total number of trials. For conciseness, we represent the 
right-hand-side term in equation~\eqref{eqn:muex} as $- k_B T\ln\langle \exp\left(-\beta \Delta \mathscr{H} \right) \rangle_{n}$ 
and here the subscript $n$ denotes that the ensemble average is taken on a membrane with $n$ proteins. The form of the excess 
chemical potential given in eqn.~\eqref{eqn:muex} has been derived by treating all insertion sites on the membrane to be 
homogeneous, while the local excess chemical potential which depends on the spatial location ${\bf x}$ on the membrane surface is 
given by, 
\begin{equation}
\mu_{P}^{ex}({\bf x}) = - k_B T\ln\langle \exp\left(-\beta\Delta \mathscr{H}({\bf x})\right) \rangle_{n}.
 \label{eqn:inhomo}
\end{equation}
In eqn.~\eqref{eqn:inhomo}, $\Delta \mathscr{H}({\bf x})$ denotes the change in energy due to the insertion of a protein field at 
spatial location ${\bf x}$. The inhomogeneous chemical potential and the spatially uniform chemical potential 
$\mu_{P}^{ex}$ can be used to determine the inhomogeneous, scaled spatial density of the proteins using the relation:
\begin{equation}
\rho({\bf x}) = \exp\left(\beta \mu_{P}\right) \exp\left(-\beta \mu_{P}^{ex}({\bf x}) \right),
 \label{eqn:inhomo-density}
\end{equation}

\subsection{Thermodynamic integration (TI) method}
Thermodynamic integration is perturbative technique that can be used to determine the relative free energy difference between any 
two thermodynamic states $A$ and $B$, provided there exists a continuous path ${\cal C}$ that connects  state $A$ to $B$; in the 
context of protein induced membrane remodeling the states $A$ and $B$ correspond to a membrane with $n$ and $n+1$ proteins 
respectively. If $\mathscr{H}_{n}$ and $\mathscr{H}_{n+1}$ be the energies of these two states then the various intermediate states 
of the membrane along the path ${\cal C}$ can be obtained by varying the coupling parameter $0 \leq \lambda \leq 1$ such that the 
energy of any intermediate state is given by $\mathscr{H}(\lambda)$, subject to the boundary condition ${\cal 
H}(0)=\mathscr{H}_{n}$ and $\mathscr{H}(1)={\cal H}_{n+1}$. The energy of an intermediate state with a given value of $\lambda$ can 
be expressed in terms of the energies of the end states as ~\cite{Frenkel:2001}:
\begin{equation}
\mathscr{H}(\lambda) = (1-\lambda)\mathscr{H}_{n} + \lambda \mathscr{H}_{n+1}.
\end{equation}

The free-energy change along this path  can expressed in its integral form as,
\begin{equation}
\Delta {\cal F}_{\rm TI} = {\cal F}_{n+1} - {\cal F}_{n} = \int_{0}^{1} \frac{\partial {\cal F}(\lambda)}{\partial \lambda} d 
\lambda.
\label{eqn:TI-pre}
\end{equation}
Using the definitions of the partition function and the free energy, given in  eqns.~\eqref{eqn:partfunc} and 
~\eqref{eqn:freeen-def}, in the above equation the relative free energy $\Delta {\cal F}_{\rm TI}$ can be shown to be

\begin{equation}
\Delta {\cal F}_{\rm TI} = \int_{0}^{1} \left\langle \frac{\partial \mathscr{H}(\lambda)}{\partial \lambda} 
\right\rangle d 
\lambda.
\label{eqn:TI}
\end{equation}

In practice, the integrand in eqn.~\eqref{eqn:TI} is estimated from independent simulations of the system at 
pre-determined, sufficiently small intervals of $\lambda$ and the relative free energy   $\Delta F_{\rm TI}$ is then 
estimated through numerical integration of eqn.~\eqref{eqn:TI}.

\subsection{Bennett acceptance ratio method (BAM)}
Bennett acceptance method is another perturbative technique that can be used to approximate the free-energy difference 
between two states close to each other in their conformational space~\cite{Bennett:1976gj}.  The microscopic 
reversibility for the transition of the membrane-protein system between the two states with $n$ and $n+1$ proteins, also 
called detailed balance condition, can be stated as
\begin{equation}
M({\mathscr{H}}_{n+1} - {\mathscr{H}}_n) \exp(-\beta {\mathscr{H}}_n) = M({\mathscr{H}}_n - {\mathscr{H}}_{n+1}) \exp(-\beta 
{\mathscr{H}}_{n+1}),
 \label{eqn:rates-equal}
\end{equation}
where $M$ is some function that defines the distribution of the acceptance probability for a transition of the membrane 
from a state with $n$ proteins to a state with $n+1$ proteins and vice versa. First, integrating both sides of 
eqn.~\eqref{eqn:rates-equal} over the entire conformational space over which the partition trace of 
eqn.~\eqref{eqn:partfunc} is defined, and then multiplying and dividing both sides with their corresponding partition 
functions the above equation can be rewritten as,

\begin{equation}
\dfrac{Q_{n} \int d \boldsymbol{\eta}\, M({\mathscr{H}}_{n+1} - {\mathscr{H}}_n) \exp(-\beta {\mathscr{H}}_n)}{Q_{n}} = 
\dfrac{Q_{n+1} \int d\boldsymbol{\eta}\,\, M({\mathscr{H}}_n - {\mathscr{H}}_{n+1}) \exp(-\beta {\mathscr{H}}_{n+1})}{Q_{n+1}},
\end{equation}
Here, for conciseness we use the state variable $\boldsymbol{\eta}$ to denote integration over all the  states of the 
triangulated surface. The above equation reduces to the form

\begin{equation}
 \frac{Q_{n+1}}{Q_n} \equiv \, \exp\left(\frac{- \Delta {\cal F}_{\rm BAM}}{k_B T}\right)_{A\rightarrow B} = 
\frac{\left\langle 
M({\cal{H}}_{n+1} - {\cal{H}}_n) \right\rangle_n} {\left\langle M({\cal{H}}_n - {\cal{H}}_{n+1}) \right\rangle_{n+1}},
\label{eqn:bennett-eqn}
\end{equation}
which gives an exact expression for the relative free energy difference denoted as $\Delta {\cal F}_{\rm BAM}$, which 
can be exactly computed if the form the the transition function $M$ is known. A common choice of  $M$ is the Metropolis 
function $M(x)=\min(1,\exp(- \beta x))$, which defines the acceptance probability according to a Boltzmann distribution. 
In eqn.~\eqref{eqn:bennett-eqn}, $\langle \cdot \rangle_{m}$ represents the ensemble average of $M$ for the transition 
from state $m \rightarrow m+1$ membrane in state $m$ to state $m^{'}$. 

\subsection{Umbrella Sampling and weighted histogram analysis method}
In a number of scenarios it is desirable to determine the statistical weight and the associated free  energy of a 
particular state of a system. Canonical simulation techniques based on equilibrium Monte Carlo or Molecular dynamics 
are not well suited for such purposes if the desired state of the system has a large energy barrier with respect to its 
equilibrium; Arrhenius' law predicts negligibly small transition rates across this energy barrier and hence one would 
require infinitely long trajectories of the system in order to generate extensive samples of the state with higher 
energy. Such at transition event is called a rare event and there are a number of techniques, such as Rosenbluth 
sampling, Wang-Landau sampling, and umbrella sampling, that can be used to access the rare states of the system within 
acceptable simulation times. In this article, we only focus on the umbrella sampling technique along with the Weighted 
Histogram Analysis Method (WHAM)~\cite{Roux:1995vi} to study the thermodynamics of large deformations in cell 
membranes.

In general, let $\boldsymbol{\zeta}$ denote an atomistic or a molecular or a continuum or a collective variable of a 
system with an equilibrium probability distribution ${\cal P}(\boldsymbol{\zeta})$ which is peaked around the 
equilibrium value $\boldsymbol{\zeta}=\boldsymbol{\zeta}^{*}$ \textemdash~conventionally the variable $\boldsymbol 
\zeta$ is called a reaction coordinate. Umbrella sampling involves the simulation of the system in ${N_{\cal B}}$ 
different windows in the presence of an additional harmonic biasing potential
\begin{equation}
{\cal B}_{i}(\boldsymbol{\zeta})=\dfrac{k_{\rm bias}}{2} \left(\zeta-\zeta_i \right)^2,
\label{eqn:Hbias}
\end{equation}
such that $\boldsymbol{\zeta}_i$ denotes the preferred value of $\boldsymbol{\zeta}$ in the $i^{\rm th}$ window and 
${\cal P}_i(\boldsymbol{\zeta})$ its probability distribution. $k_{\rm bias}$ is the strength of the biasing spring 
which is chosen such that the probability distributions from neighboring windows show considerable overlap. The 
probability distributions ${\cal P}_i(\boldsymbol{\zeta})$ computed across multiple simulations windows can be combined 
together using the Weighted Histogram Analysis method to estimate the free energies of all intermediate states, with 
respect to the first window.  The free energy is computed by self-consistently solving the two WHAM equations, for the 
unknowns ${\cal P}(\boldsymbol{\zeta})$ and  ${\cal F}_i$, given by: \\
\begin{equation}
 {\cal P}(\boldsymbol{\zeta})=\dfrac{\sum\limits_{i=1}^{N_{\cal B}} {\cal N}_{i} {\cal 
P}_i(\boldsymbol{\zeta)}}{\sum\limits_{i=1}^{N_{\cal B}}{\cal N}_i \exp\left(-\beta \left({\cal 
B}_i(\boldsymbol{\zeta})-{\cal F}_i  \right) \right)},
\end{equation}
where ${\cal N}_i$ is the number of samples in the $i^{\rm th}$ window and 

\begin{equation}
 {\cal F}_i = -k_BT \ln \left( \sum\limits_{j=1}^{nbins} {\cal P}(\boldsymbol{\zeta}_j)  \exp\left( -\beta {\cal 
B}_i(\boldsymbol{\zeta}_j)\right) \right)+C.
 \label{eqn:wham-freeenergy}
\end{equation}
Here $nbins$ denotes the number of used over which the free energy is discretized and $C$ is an arbitrary constant.
\section{Predicting transition from curvature sensing to curvature inducing behavior using Widom insertion}
An important question in the area of protein driven curvature remodeling of membranes is ``when does a cluster of 
proteins behave in a cooperative manner?'' {\it In vitro}  experiments on a number of curvature inducing proteins such 
as BAR domains, ENTH domains, and Exo70 domains have shown that these proteins when at low concentrations localize to 
high curvature regions on the membrane generated by thermal undulations \textemdash~ commonly known as the {\em 
curvature sensing behavior} \textemdash~ while at high concentrations they aggregate into clusters and spontaneously 
generate membrane curvature to form highly curved membrane morphologies such as tubules and blebs \textemdash~ a 
characteristic of {\em curvature inducing behavior}.  Delineating this transition regime is a challenge in experiments 
but computational models based on free energy methods are well suited for this purpose. The relative energy free energy 
of the membrane is an excellent marker for the curvature sensing to transition behavior. It was pointed out in  
Sec.~\ref{sec:overview} that the relative free energy to introduce the $n+1^{\rm th}$ protein in a membrane with $n$ 
proteins can be computed using Widom insertion or thermodynamic integration or Bennett-acceptance-method. Here we use 
the computationally less expensive Widom insertion technique to determine the relative free energies to insert a protein 
on a membrane with $n=0$ (i.e. a pure lipid membrane).

\begin{figure}[!h]
\centering
\includegraphics[width=15cm,clip]{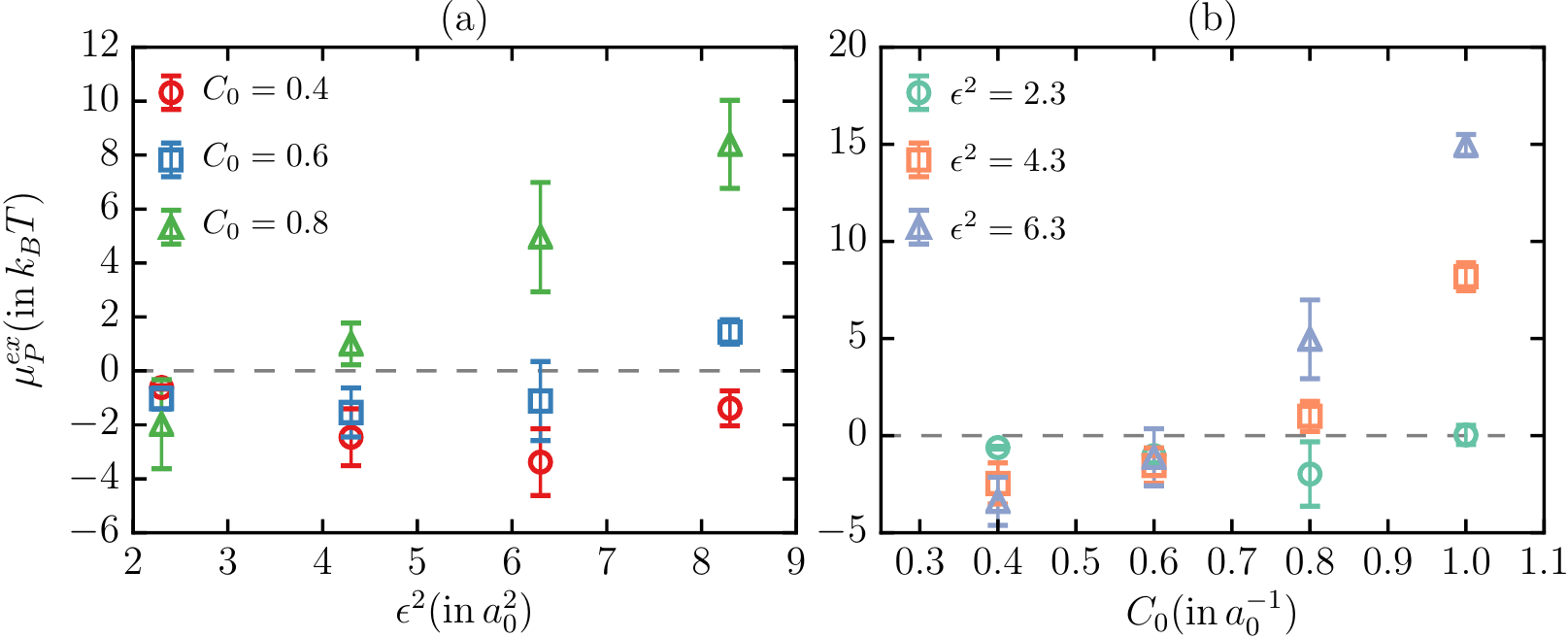}
\caption{\label{fig:muexcess} Excess chemical potential, in units of $k_BT$, to insert a protein field with maximum 
spontaneous curvature $C_0$ and extent of curvature $\epsilon^2$ on a membrane with zero proteins \textemdash~ both 
$C_0$ and $\epsilon^2$ are expressed in units of $a_0$. (a) $\mu_P^{ex}$ as a function of $\epsilon^2$ for fixed values 
of $C_0=0.4a_0^{-1}$, $0.6a_0^{-1}$, and $0.8a_0^{-1}$ and (b) $\mu_P^{ex}$ as a function of $C_0$ for fixed values of 
$\epsilon^2=2.3a_0^2$, $4.3a_0^{2}$, and $6.3a_0^{2}$}
\end{figure}

In the continuum description a protein is represented as a curvature field with a Gaussian profile that is parameterized 
using two variables namely  the maximum spontaneous curvature ($C_0$) and the extent of the curvature field 
($\epsilon^2$), see eqns.~\eqref{eqn:czeroform} and ~\eqref{eqn:curvprofile}. We express both $C_0$ and $\epsilon^2$ in 
units of $a_0$, which represents the hard sphere diameter associated with a vertex of the triangulated surface. The 
excess chemical potentials $\mu_P^{ex}$ as a function of $\epsilon^2$ (for fixed values of $C_0$) and $C_0$ (for fixed 
values of $\epsilon^2$) are shown in Figs.~\ref{fig:muexcess}(a) and ~\ref{fig:muexcess}(b) respectively. It can be seen 
that $\mu_P^{ex}$ is negative for small values of $C_0$ and $\epsilon^2$ which indicates that the free energy of the 
system is reduced upon introduction of the protein. Assuming the entropic contribution to be negligible, this implies 
that the total bending potential given by eqn.~\eqref{eqn:can-Helf-discrete} is smaller in the presence of the protein 
which is only possible when $H \approx C_0$. Since the Widom test particle only probes the membrane curvature and does 
not deforms the membrane it is clear that the equilibrium curvature profile of the membrane matches that of the inserted 
protein field which is characteristic of curvature sensing behavior.  On the other hand when $C_0$ and $\epsilon^2$ are 
larger ($C_0>H$)  $\mu^{ex}_P$ become large and positive and since such states are thermodynamically unstable any such 
proteins associated with the membrane would tend to to generate local curvatures that match their intrinsic curvature 
profile and this regime where the protein induces curvature. The results presented in Fig.~\ref{fig:muexcess} only 
focus on the thermodynamic stability of a single protein but even weakly curving proteins can transition from curvature 
sensing to inducing behavior, when their concentration exceeds a critical value. The effect of the cooperative 
behavior, due to the self- and membrane-mediated interactions of the proteins, on the curvature 
inducing properties of membrane associated proteins  has been recently studied in the context of ENTH 
domains~\cite{Tourdot:2014wh,Bradley:2014hm}.

\section{Comparing predictions from Widom insertion, TI, and BAM}
The excess chemical potential is reflective of the underlying free energy landscape as shown in Fig.~\ref{fig:muexcess} 
and in section we compare these predictions to the corresponding relative free energy levels obtained using 
thermodynamic integration and Bennett-acceptance-method. The free energies determined using TI and BAM (see 
eqns.~\eqref{eqn:TI} and ~\eqref{eqn:bennett-eqn}) are related to the total chemical potential 
$\mu_P=\mu_{P}^{id}(\rho)+\mu_P^{ex}$ as:
\begin{equation}
\Delta F_{\rm TI} = \Delta F_{\rm BAM} =  \mu_{P}.
\label{eqn:compare-ti-b-w}
\end{equation}
Since the Widom insertion technique can only be used to determine the excess part of the chemical potential the various 
free energies can be compared only if the total chemical potential can be determined. $\mu_{P}^{id}(\rho)$ is 
the entropic configurational component of $\mu_P$ and depends on the number of conformational states visited by a 
single particle or protein-field. The number of conformational states accessible to the $n+1^{\rm th}$ protein on a 
membrane with $n$ proteins can in turn be determined using the trajectories obtained using TI. In TI, the additional 
protein field  is grown from a non-existent entity to a full-existent object by varying the parameter $\lambda$ from $0$ 
to $1$. The degree of localization changes with change in $\lambda$ and when $\lambda \rightarrow 1$ the protein does 
not explore all the conformational states but is instead confined to the minimum of the free energy well, and this 
minimum in the triangulated surface model corresponds to a subset of vertices on the surface.  The required  
correction $\mu_{P}^{id}(\rho)$ can be calculated from the number of unique vertices $N_{\psi}$ visited by a 
protein field in various TI simulations with $\lambda \sim 1$ as:
\begin{equation}
\mu_{P}^{id}(\rho) = - k_B T \ln \left( \frac{2 \sigma_{\psi}}{N} \right).
\label{eqn:widom-correction}
\end{equation}
Here $N$ is the total number of vertices on the triangulated surface and the standard deviation $\sigma_{\psi}$ 
of the distribution of unique vertices is computed as, 
\begin{equation}
\sigma_{\psi} = \left( \sum\limits_{\upsilon=1}^{N_\psi} \upsilon^2 P_{\upsilon} - 
\left(\sum\limits_{\upsilon=1}^{N_\psi}  \upsilon P_{\upsilon} \right)^2 \right )^{\dfrac{1}{2}},
\label{eqn:variance-correction}
\end{equation}
with $P_{\upsilon}$ being the probability of a protein to visit the $\upsilon^{\rm th}$ unique vertex.

\begin{figure}[!h]
\centering
\includegraphics[width=15cm,clip]{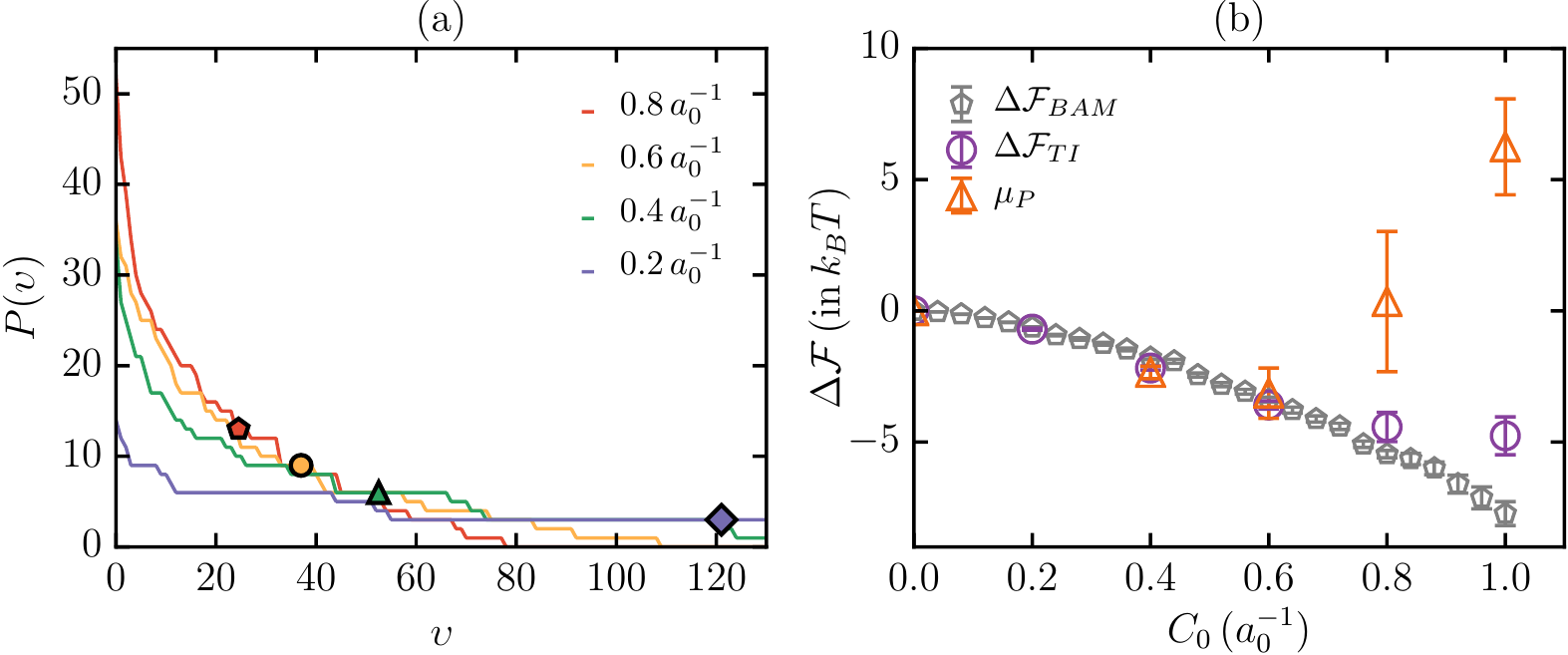}
\caption{\label{fig:distribution} (a) Distribution of the number of unique vertices $P(\upsilon)$, visited in a TI 
simulation with $\lambda \sim 1$, for four different values of $C_0$. The points shown alongside each curve correspond 
to the standard deviation $\sigma_{\psi}$. (b) Comparison of the relative free energies to add one protein to a 
membrane with zero proteins computed using TI, BAM, and Widom insertion.}
\end{figure}

The probability distribution of the number of unique vertices, for four different values of $C_0$,  is shown in 
Fig.~\ref{fig:distribution}(a) and the symbols shown alongside correspond to the value of $\sigma_\psi$. The total 
chemical potential that compares the entropic configurational part is compared against the relative free energies 
computed using TI and BAM in Fig.~\ref{fig:distribution}(b) . It can be seen from Fig.~\ref{fig:distribution}(b) that 
all three methods agree well for low values of $C_0$ while Widom insertion deviates from the other methods above $C_0 > 
0.6 a_0^{-1}$. The deviation of TI and Widom insertion methods at high $C_0$ is well known since efficient sampling of 
$\mu_P^{ex}$ suffers for large perturbations in energy or higher densities.

\section{In silico tether pulling experiments}
\begin{figure}[!h]
\centering
\includegraphics[width=12.5cm,clip]{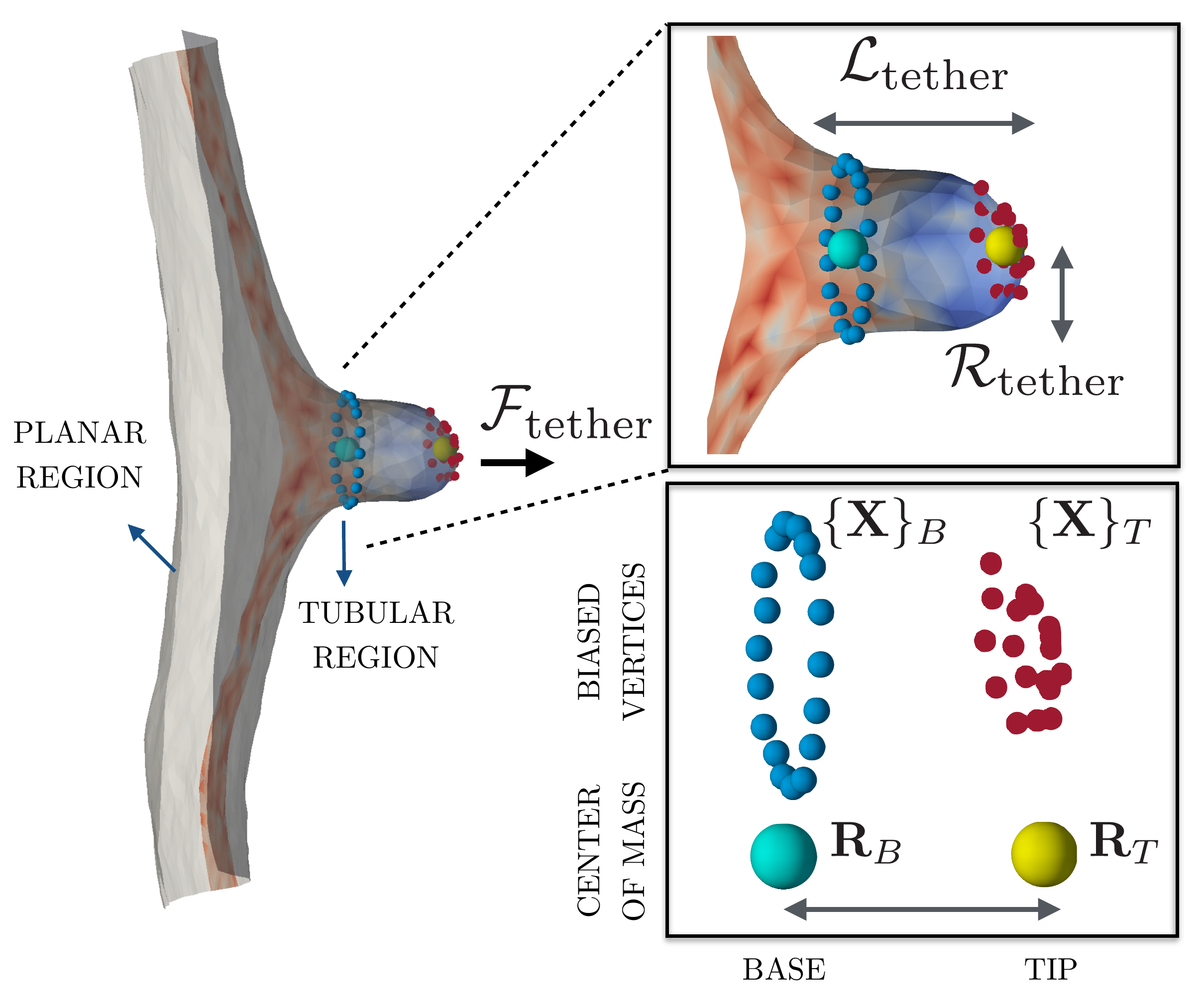}
 \caption{\label{fig:tether-illus} (left panel) A snapshot of a tether extracted from a patch of a  planar membrane. 
(right panel) A closer view of the tubular region of length \lent{} and radius \radt{} with the colors denoting the 
mean curvature of the surface \textemdash~ the tubular region has a positive mean curvature and the neck region has a 
negative mean curvature. The tip and base regions on the membrane tether are represented by the marked vertices on the 
tubular membrane along with the position of their center of mass.}
\end{figure}

Extraction of cylindrical protrusions (tethers) from the surface of a cell membrane, using optical tweezers or 
functionalized AFM tips or through attachment of magnetic beads,  is a useful method to characterize its mechanical 
properties such as the bending stiffness, surface tension, and degree of cytoskeletal pinning. A tether is characterized 
by its radius \radt{}, its length \lent{} and the force required for its extraction denoted by \fort{}, as shown in the 
illustration in Fig.~\ref{fig:tether-illus}. In order to clearly delineate the role of the various parameters 
characterizing a cell membrane in a typical {\it in vivo} tether extraction assay, it is essential to develop physical 
models that allow us to gain an understanding at a fundamental level. 

In this section, we present an {\it in silico} tether extraction assay by combining the triangulated surface membrane 
with umbrella sampling techniques and the weighted histogram analysis method. In order to stabilize a membrane tether of 
length \lent{}, we apply a  umbrella sampling biasing potential on a set of macroscopic variables which are defined as 
follows. The tip of the tubular region is represented by a set of pre-determined vertices  $\{{\bf X}\}_T$ with center 
of mass ${\bf R}_T$ and the base of the tether is represented by another set of of vertices $\{{\bf X}\}_B$ with center 
of mass ${\bf R}_B$,  such that  ${\cal L}_{\rm tether}=|{\bf R}_T-{\bf R}_B|$ and  each vertex  in $\{{\bf X}\}_B$ with 
position vector ${\bf x}_B$ obeys the constraint $|{\bf R}_T-{\bf x}_B| \leq 1.5$\lent{}. The macroscopic positions of 
the tip and base of the membrane tether ${\bf R}_T$ and ${\bf R}_B$ define an order parameter which is subjected to a 
harmonic biasing potential in the $n^{\rm th}$ window given as:
\begin{equation}
{\cal B}_{n}({\bf R}_T, {\bf R}_B) = \dfrac{k_{\rm bias}}{2} \left(\left|{\bf R}_T-{\bf R}_B \right|-{\cal L}_{\rm tether}^{*} \right)^2
\end{equation}

\begin{figure}[!h]
\centering
\includegraphics[width=12.5cm,clip]{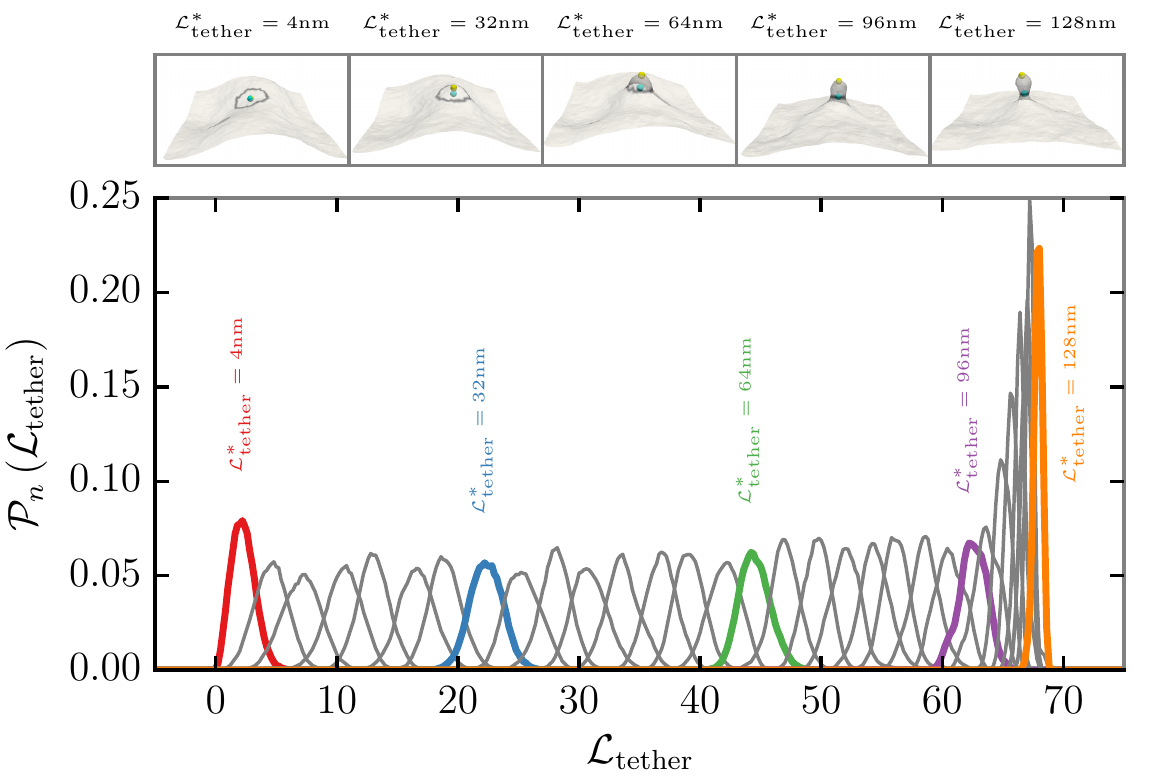}
 \caption{\label{fig:tether-histogram} (top panel) Snapshots of a membrane subject to a biasing potential, along with 
the positions of the center of masses, for five different values of the preferred tether length ${\cal L}_{\rm 
tether}^{*}=$ 4, 32, 64, 96, and 128  nm. (bottom panel) ${\cal P}_{n}({\cal L}_{\rm tether})$, the probability 
distribution of the tether length ${\cal L}_{\rm tether}$ in 32 different biasing windows with a window size of 4 nm. 
The curve corresponding to the snapshots in the top panel are shown as solid colored lines along with the corresponding 
values of the preferred tether length.}
\end{figure}

$k_{\rm bias}$ is the strength of the biasing potential and ${\cal  L}_{\rm tether}^{*}$ denotes the preferred tether 
length. The conformational state of the membrane patch is evolved using the Dynamical Triangulation Monte Carlo 
technique with the total potential $\mathscr{H}_{\rm tot}=\mathscr{H}+{\cal B}_n$. The conformations of a membrane, with 
fixed values of $\kappa=20k_{B}T$, ${\cal L}=510{\rm nm}$, and $A_{ex}=10\%$, in five different biasing windows with 
${\cal L}_{\rm tether}^{*}=$ 4, 32, 64, 96, and 128 nm are shown in the top panel of Fig.~\ref{fig:tether-histogram}. 
The positions of the center of masses corresponding to the biasing vertices, ${\bf R}_T$ and ${\bf R}_B$ respectively, 
are also shown alongside and it can be seen that tether like structures are readily formed  at larger values of ${\cal 
L}_{\rm tether}^{*}$. 

The probability distribution of the tether length \lent{} in 32 different sampling windows, with a window size of 4 nm, 
are shown in the bottom panel of Fig.~\ref{fig:tether-histogram}. ${\cal P}_{n}({\cal L}_{\rm tether})$  shows a normal 
distribution in each of the 32 windows and the peak of the distribution shifts to a higher value of ${\cal L}_{\rm 
tether}$ with increasing  ${\cal L}_{\rm tether}^{*}$ and the strength of the biasing potential $k_{\rm bias}$ was 
chosen so that distributions from adjacent windows show a good overlap as seen in  Fig.~\ref{fig:tether-histogram}. It 
can also be seen that ${\cal P}_{n}({\cal L}_{\rm tether})$ becomes narrower following the formation of the tether at 
${\cal L}_{\rm tether}^{*} \approx 96{\rm nm}$.

\begin{figure}[!h]
\centering
\includegraphics[width=12.5cm,clip]{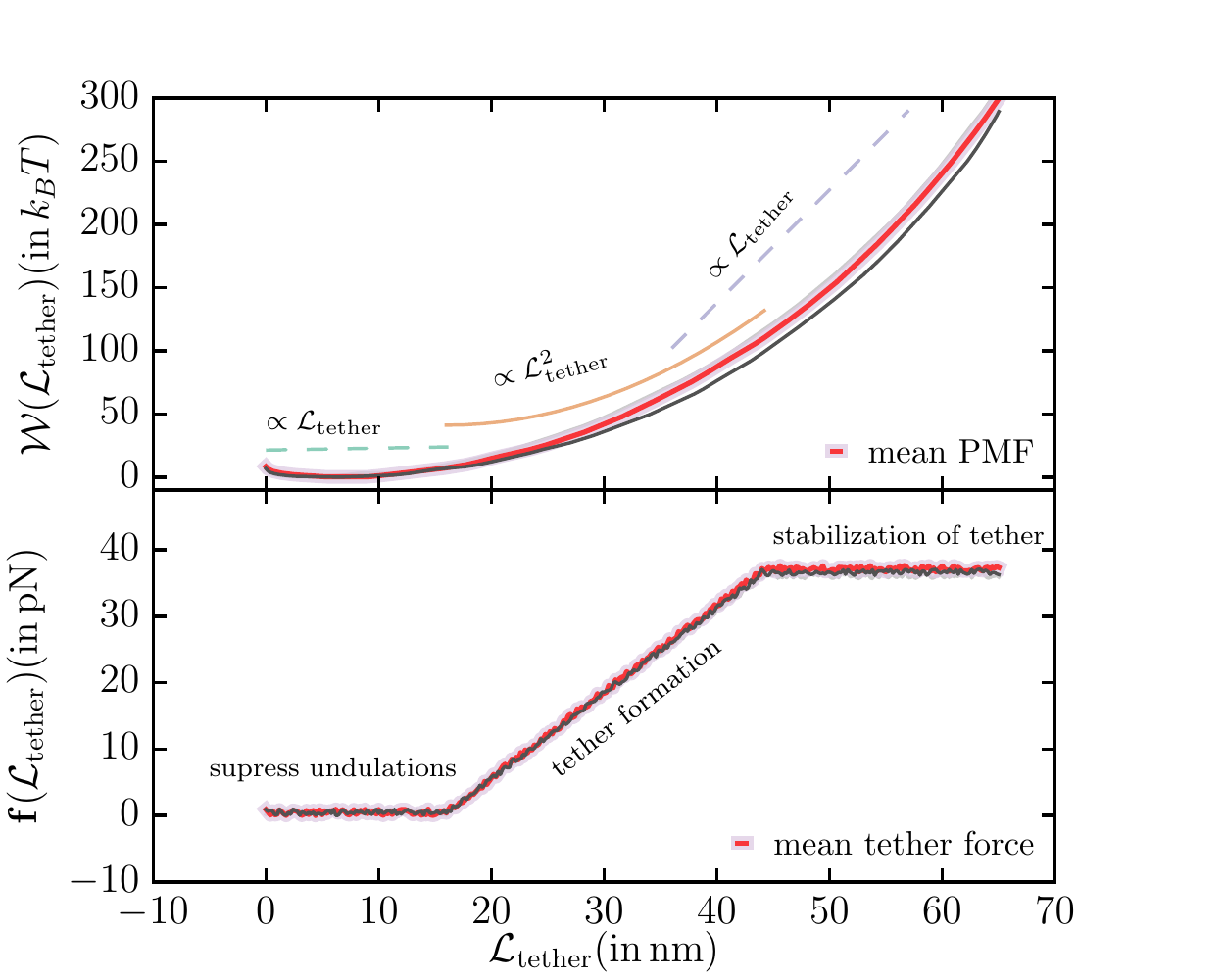}
 \caption{\label{fig:tether-pmf-force} (top panel) Potential of mean force  (PMF) ${\cal W}_{\rm tether}$, in units of 
$k_{B}T$, as a function of the tether length ${\cal L}_{\rm tether}$. The PMF shows three distinct scaling regimes 
\textemdash~an initial linear regime followed by a quadratic regime which crosses over to a final linear regime 
\textemdash~ and the scaling relations are also shown alongside.  (bottom panel) The force ${\cal F}_{\rm tether}$, in 
units of pN, required to extract the tether.} 
\end{figure}

The potential of mean force ${\cal W}({\cal L}_{\rm tether})$ which denotes the energy required to extract a tether of 
length ${\cal L}_{\rm tether}$, computed by combining the histograms in Fig.~\ref{fig:tether-histogram} using WHAM, is 
shown in the top panel of Fig.~\ref{fig:tether-pmf-force}. The PMF shows three distinct regimes which are also shown 
alongside the PMF in Fig.~\ref{fig:tether-pmf-force}: (i) an initial weakly linear regime ($\propto {\cal L}_{\rm 
tether}$), (ii) an intermediate quadratic regime ($\propto {\cal L}_{\rm tether}^2$), and (iii) a final linear regime 
($\propto {\cal L}_{\rm tether})$. These three regimes have a significance in the formation and stabilization of the 
membrane tether. When a force is applied to an undulating membrane the short wavelength undulations in the membrane 
conformation are suppressed in the linear response regime and this response characterizes the initial linear regime. The 
tubular structures nucleate and grow in the intermediate quadratic regime  until all the undulations in the membrane are 
ironed out and are drawn into the tubular region \textemdash~ the extent of the quadratic regime changes with change in 
the membrane excess area which sets the intensities of the characteristic undulations in the 
membrane~\cite{Ramakrishnan:ureview}. In the final linear regime the tether does not extends considerably and all the 
applied force (i.e. the biasing potential in the current context) is primarily used to stabilize the length of membrane 
tether and this leads to reduced undulations in the tether length which is seen in the narrow distribution of \lent{} at 
large values of ${\cal L}_{\rm tether}^{*}$. The force required to extract a tether can be determined from the PMF as 
${\bf f}({\cal L}_{\rm tether})=-\nabla_{{\cal L}_{\rm tether}}{\cal W}$. Numerical differentiation of the PMF can lead 
to large errors in the estimates for the tether force and hence we use an alternate method where we utilize the scaling 
relations to determine ${\bf f}({\cal L}_{\rm tether})$. The tether force determined using the scaling relations are 
shown in the bottom panel of Fig.~\ref{fig:tether-pmf-force} and the constant force in the final regime is taken to be 
the tether force that can be compared to that obtained in tether pulling experiments. Estimates for force and the radius 
of the membrane tether computed using the continuum {\it in silico} assay described here are in excellent agreement with 
those reported in the literature for cells with similar mechanical properties, and this assay has been used by 
Ramakrishnan et al. to determine the excess area in membrane regions between cytoskeletal pinning points. 

\section{Bridging techniques using WHAM and TI}
The PMF of a system determined using WHAM analysis is accurate only upto an additive constant, as shown in 
eqn.~\eqref{eqn:wham-freeenergy}, and hence can only be used to determine the relative energy differences between the 
various states. Furthermore, the umbrella sampling technique and WHAM also suffer from the problem of {\it 
end-point-catastrophe}, which is a well known phenomenon in alchemical energy methods commonly used to study the free 
energy landscapes of biomolecular systems. In such systems, the free energy estimates close to the end points of the 
order parameter are erroneous due to the numerical instabilities arising from the divergence of the interaction 
potential \textemdash~ a case point being the free energy calculation involving two atoms, interacting via a 
Lennard-Jones potential, with their separation $r \rightarrow 0$. WHAM also suffers from such shortcomings at the end 
points albeit for a different reason. An end point is defined as the value of the order parameter beyond which the 
potential of mean force vanishes since the various underlying interaction potentials vanish. In this regime, where the 
strength of the PMF is very weak (compared to the strength of the fluctuations), the signal to noise ratio is very small 
and hence conventional sampling techniques do not yield the correct probability distribution for the order parameter in 
the windows close to the end point. As a result the PMF computed using WHAM does not accurately capture the energy 
landscape of the system close to the end point. The problem of end-point-catastrophe can be overcome either by 
generating infinitely long samples or by performing the simulations using a stronger biasing potential with much smaller 
window sizes and both these methods leads to large computational costs. A similar method has been discussed in the 
context of solvation free energies by Souaille and Roux~\cite{Souaille:2001gm}.
 
 In certain class of problems, the thermodynamic integration technique can be used to overcome the end-point-catastrophe 
mentioned above and also fix the absolute energy levels for the PMF. We demonstrate the idea of bridging the free 
energies computed using WHAM and TI using the case of a functionalized nanocarrier interacting with receptor molecules 
expressed on the surface of  a flat membrane. We follow the model previously described by Liu et 
al~\cite{Liu:2010em,Liu:2011fe} to construct the free energy landscape for the interaction of an anti-ICAM 
functionalized nanocarrier  with a membrane surface expressing ICAM receptors. In brief, the nanocarrier is a sphere of 
radius 50 nm, discretized into 162 vertices, and is functionalized with antibodies, that are represented as radial 
vectors of length 15 nm. The membrane is represented as a planar substrate on which the receptors molecules are modeled 
as cylinders of length 19 nm and radius 1.5 nm. The interaction between the tip of an antibody and the tip of a surface 
receptor is modeled as a Bell bond with a potential:
 \begin{equation}
 G(d_{ij}) = G_{0}+\dfrac{k}{2} d_{ij}^{2}.
 \end{equation}
$G_{0}$ is the activation energy gained by the system when a bond is formed, $k$ is the strength of receptor-antibody 
bond, $d_{ij}$ is the distance between the tips of the antibody and receptor molecules. $d_0$ denotes the maximum 
extension of the antibody-receptor bonds and the bonds break when $d_{ij}>d_{0}$. Typical values for 
the anti ICAM-ICAM interactions have been shown to be $G_{0}=-19.4k_{B}T$, $k=1$N/m, $d_{0} \sim 0.4$ nm, and the 
receptor density being  2000 ICAM/$\mu$m$^{2}$~\cite{Liu:2010em}. 

\begin{figure}[!h]
\centering
\includegraphics[width=15cm,clip]{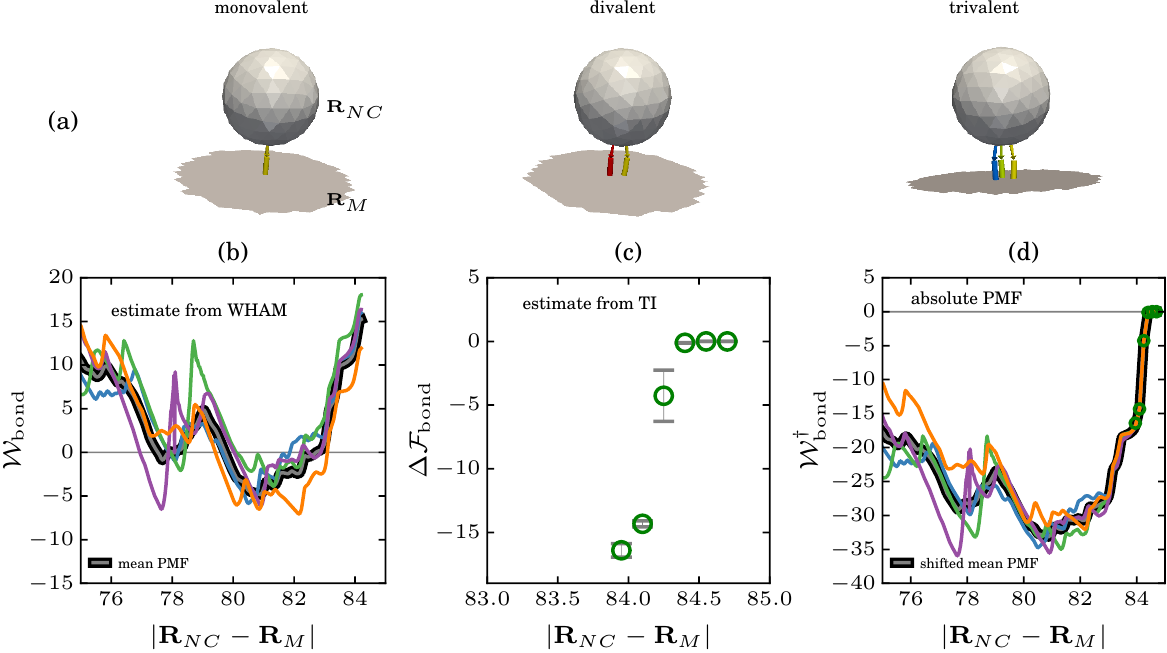}
 \caption{\label{fig:WHAM-TI} (a) Snapshots of a functionalized nanocarrier forming one, two, and three simultaneous 
bonds with the receptor molecules expressed on the membrane surface. ${\bf R}_{NC}$ and ${\bf R}_{M}$ denote the center 
of mass position of the nanocarrier and membrane respectively, (b) The potential of mean force ${\cal W}_{\rm bond}$ as 
a function of $|{\bf R}_{NC}$-${\bf R}_{M}|$ for four different ensembles computed using umbrella sampling and WHAM, (c) 
the relative free energy difference  $\Delta {\cal F}_{\rm bond}$ between a nanocarrier with zero and one 
antigen-antibody bonds, and (d) the absolute free energy of the system obtained by combining the PMF obtained using 
WHAM the relative free energy obtained using TI.} 
\end{figure}

A functionalized nanocarrier can form multiple simultaneous antibody-receptor bonds and the degree of bonding varies 
with position of the nanocarrier (${\bf R}_{NC}$) with respect to the base of the planar membrane (${\bf R}_{M}$) and 
hence the relative distance $|{\bf R}_{NC}-{\bf R}_{M}|$ is a suitable choice for the reaction coordinate along the 
potential of mean force is to be evaluated. For the system considered here, it should be noted that the nanocarrier 
cannot form bonds when $|{\bf R}_{NC}-{\bf R}_{M}|>$84.4 nm\footnote{calculated as (nanocarrier radius + length of the 
antibody + length of the receptor + $d_{0}$)} and hence the potential of mean force is zero beyond this value of the 
reaction coordinate. Hence we take the region around $|{\bf R}_{NC}-{\bf R}_{M}|=$84.4 nm to represent the regime 
corresponding to the end point catastrophe. 

The conformations of a nanocarrier with one, two, and three simultaneous bonds  obtained at different values of the 
reaction coordinates are shown in Fig.~\ref{fig:WHAM-TI}(a) (the unbound antibodies and receptors are not shown for 
clarity).  ${\cal W}_{\rm bond}$, the PMF computed using the umbrella sampling/WHAM techniques, through the application 
of a biasing potential on the reaction coordinate, is shown in Fig.~\ref{fig:WHAM-TI}(b) for four different ensembles  
of the nanocarrier-membrane system. It can be seen that  the potential of mean force obtained from different ensembles 
are shifted upto an arbitrary constant but the relative energy levels between the various states remain nearly constant, 
which is reflective of the first shortcoming discussed at the introduction. A closer look at the region between 84.0 and 
84.4 nm reveals that the harmonic potential, characteristic of a single antibody-receptor bond with a well depth of 
19$k_{B}T$, is only partially captured in the WHAM analysis and this is a clear signature of the end-point catastrophe.

On the other hand, as shown in Fig.~\ref{fig:WHAM-TI}(c), the free energy difference close to the end point region 
computed using TI coupled with WHAM precisely captures both the non-bonded region ($|{\bf R}_{NC}-{\bf R}_{M}|>$84.4 nm) 
and also the depth of potential well corresponding to a single bond. These results were obtained using short TI 
calculations performed in the end point region with an window interval of 0.1 nm. Since the bonded and unbonded states 
are clearly defined the relative free energy computed using TI can be combined with PMF computed using WHAM to fix the 
end-point-catastrophe and also represent the PMF in an absolute scale. The absolute potential of mean force (${\cal 
W}_{\rm bond}^{\dagger}$) obtained by shifting ${\cal W}_{\rm bond}$  with respect to $\Delta {\cal F}_{\rm bond}$ 
through a linear regression fit is shown in Fig.~\ref{fig:WHAM-TI}(d). The absolute PMF obtained by combining WHAM and 
TI is in excellent agreement with the reported values of the PMF obtained by grafting the end points using an analytic 
function~\cite{Liu:2010em}. The bridging technique presented here is very generic and can be applied to even more 
complex scenarios where the exact form of the analytic function as a function of the order parameter is not known.

\section{Conclusions}
The analysis of the thermodynamic free energy landscape can yield unprecedented levels of insight into the behavior of 
complex systems. However, it is a challenge in the study of these systems to formulate computational methods to 
delineate their free energy landscape using physically relevant order parameters. The primary focus of this article is 
to demonstrate how conventional free energy methods can be adopted to in problems related to morphological transitions 
in cell membrane.  Quantitative predictions based on the relative free energies obtained using these simple but elegant 
methods reproduce many of the emergent behaviors observed in  experiments. Since these methods provide a powerful 
framework to interpret experimental findings it is essential to develop free energy based models and methods which 
help in understanding the system at a more fundamental level. 

\section*{Acknowledgments}
This work was supported in part by the US National Science Foundation Grants DMR-1120901, and CBET-1244507. The research leading to these results has received funding from the European Commission Grant FP7-ICT-2011-9-600841, US NIH U01-EB016027, and NIH 1U54CA193417. Computational resources were provided in part by the National Partnership for Advanced Computational Infrastructure under Grant No. MCB060006 from XSEDE.

\bibliographystyle{aipnum4-1}

\begin{thebibliography}{45}%
	\makeatletter
	\providecommand \@ifxundefined [1]{%
		\@ifx{#1\undefined}
	}%
	\providecommand \@ifnum [1]{%
		\ifnum #1\expandafter \@firstoftwo
		\else \expandafter \@secondoftwo
		\fi
	}%
	\providecommand \@ifx [1]{%
		\ifx #1\expandafter \@firstoftwo
		\else \expandafter \@secondoftwo
		\fi
	}%
	\providecommand \natexlab [1]{#1}%
	\providecommand \enquote  [1]{``#1''}%
	\providecommand \bibnamefont  [1]{#1}%
	\providecommand \bibfnamefont [1]{#1}%
	\providecommand \citenamefont [1]{#1}%
	\providecommand \href@noop [0]{\@secondoftwo}%
	\providecommand \href [0]{\begingroup \@sanitize@url \@href}%
	\providecommand \@href[1]{\@@startlink{#1}\@@href}%
	\providecommand \@@href[1]{\endgroup#1\@@endlink}%
	\providecommand \@sanitize@url [0]{\catcode `\\12\catcode `\$12\catcode
		`\&12\catcode `\#12\catcode `\^12\catcode `\_12\catcode `\%12\relax}%
	\providecommand \@@startlink[1]{}%
	\providecommand \@@endlink[0]{}%
	\providecommand \url  [0]{\begingroup\@sanitize@url \@url }%
	\providecommand \@url [1]{\endgroup\@href {#1}{\urlprefix }}%
	\providecommand \urlprefix  [0]{URL }%
	\providecommand \Eprint [0]{\href }%
	\providecommand \doibase [0]{http://dx.doi.org/}%
	\providecommand \selectlanguage [0]{\@gobble}%
	\providecommand \bibinfo  [0]{\@secondoftwo}%
	\providecommand \bibfield  [0]{\@secondoftwo}%
	\providecommand \translation [1]{[#1]}%
	\providecommand \BibitemOpen [0]{}%
	\providecommand \bibitemStop [0]{}%
	\providecommand \bibitemNoStop [0]{.\EOS\space}%
	\providecommand \EOS [0]{\spacefactor3000\relax}%
	\providecommand \BibitemShut  [1]{\csname bibitem#1\endcsname}%
	\let\auto@bib@innerbib\@empty
	\bibitem [{\citenamefont {Israelachvili}(2011)}]{Israelachvili:2011wy}%
	\BibitemOpen
	\bibfield  {author} {\bibinfo {author} {\bibfnamefont {J.~N.}\ \bibnamefont
			{Israelachvili}},\ }\href@noop {} {\emph {\bibinfo {title} {{Intermolecular
					and Surface Forces}}}},\ \bibinfo {edition} {third edition}\ ed.\ (\bibinfo
	{publisher} {Academic Press},\ \bibinfo {address} {Boston},\ \bibinfo {year}
	{2011})\BibitemShut {NoStop}%
	\bibitem [{\citenamefont {Escrib{\'a}}\ \emph {et~al.}(2008)\citenamefont
		{Escrib{\'a}}, \citenamefont {Gonz{\'a}lez-Ros}, \citenamefont {Go{\~n}i},
		\citenamefont {Kinnunen}, \citenamefont {Vigh}, \citenamefont
		{S{\'a}nchez-Magraner}, \citenamefont {Fern{\'a}ndez}, \citenamefont
		{Busquets}, \citenamefont {Horv{\'a}th},\ and\ \citenamefont
		{Barcel{\'o}-Coblijn}}]{Escriba:2008hb}%
	\BibitemOpen
	\bibfield  {author} {\bibinfo {author} {\bibfnamefont {P.~V.}\ \bibnamefont
			{Escrib{\'a}}}, \bibinfo {author} {\bibfnamefont {J.~M.}\ \bibnamefont
			{Gonz{\'a}lez-Ros}}, \bibinfo {author} {\bibfnamefont {F.~M.}\ \bibnamefont
			{Go{\~n}i}}, \bibinfo {author} {\bibfnamefont {P.~K.~J.}\ \bibnamefont
			{Kinnunen}}, \bibinfo {author} {\bibfnamefont {L.}~\bibnamefont {Vigh}},
		\bibinfo {author} {\bibfnamefont {L.}~\bibnamefont {S{\'a}nchez-Magraner}},
		\bibinfo {author} {\bibfnamefont {A.~M.}\ \bibnamefont {Fern{\'a}ndez}},
		\bibinfo {author} {\bibfnamefont {X.}~\bibnamefont {Busquets}}, \bibinfo
		{author} {\bibfnamefont {I.}~\bibnamefont {Horv{\'a}th}}, \ and\ \bibinfo
		{author} {\bibfnamefont {G.}~\bibnamefont {Barcel{\'o}-Coblijn}},\ }\href
	{\doibase 10.1111/j.1582-4934.2008.00281.x} {\bibfield  {journal} {\bibinfo
			{journal} {J Cellular Mol Med}\ }\textbf {\bibinfo {volume} {12}},\ \bibinfo
		{pages} {829} (\bibinfo {year} {2008})}\BibitemShut {NoStop}%
	\bibitem [{\citenamefont {Singer}\ and\ \citenamefont
		{Nicolson}(1972)}]{Singer:1972uq}%
	\BibitemOpen
	\bibfield  {author} {\bibinfo {author} {\bibfnamefont {S.~J.}\ \bibnamefont
			{Singer}}\ and\ \bibinfo {author} {\bibfnamefont {G.~L.}\ \bibnamefont
			{Nicolson}},\ }\href
	{http://adsabs.harvard.edu/cgi-bin/nph-data_query?bibcode=1972Sci...175..720S&link_type=EJOURNAL}
	{\bibfield  {journal} {\bibinfo  {journal} {Science}\ }\textbf {\bibinfo
			{volume} {175}},\ \bibinfo {pages} {720} (\bibinfo {year}
		{1972})}\BibitemShut {NoStop}%
	\bibitem [{\citenamefont {Edidin}(2003)}]{Edidin:2003fv}%
	\BibitemOpen
	\bibfield  {author} {\bibinfo {author} {\bibfnamefont {M.}~\bibnamefont
			{Edidin}},\ }\href {\doibase 10.1038/nrm1102} {\bibfield  {journal} {\bibinfo
			{journal} {Nat. Rev. Mol. Cell Biol.}\ }\textbf {\bibinfo {volume} {4}},\
		\bibinfo {pages} {414} (\bibinfo {year} {2003})}\BibitemShut {NoStop}%
	\bibitem [{\citenamefont {Engelman}(2005)}]{Engelman:2005bo}%
	\BibitemOpen
	\bibfield  {author} {\bibinfo {author} {\bibfnamefont {D.~M.}\ \bibnamefont
			{Engelman}},\ }\href {\doibase 10.1038/nature04394} {\bibfield  {journal}
		{\bibinfo  {journal} {Nature Cell Biology}\ }\textbf {\bibinfo {volume}
			{438}},\ \bibinfo {pages} {578} (\bibinfo {year} {2005})}\BibitemShut
	{NoStop}%
	\bibitem [{\citenamefont {Conner}\ and\ \citenamefont
		{Schmid}(2003)}]{Conner:2003gm}%
	\BibitemOpen
	\bibfield  {author} {\bibinfo {author} {\bibfnamefont {S.~D.}\ \bibnamefont
			{Conner}}\ and\ \bibinfo {author} {\bibfnamefont {S.~L.}\ \bibnamefont
			{Schmid}},\ }\href {\doibase 10.1038/nature01451} {\bibfield  {journal}
		{\bibinfo  {journal} {Nature}\ }\textbf {\bibinfo {volume} {422}},\ \bibinfo
		{pages} {37} (\bibinfo {year} {2003})}\BibitemShut {NoStop}%
	\bibitem [{\citenamefont {Doherty}\ and\ \citenamefont
		{McMahon}(2009)}]{Doherty:2009fz}%
	\BibitemOpen
	\bibfield  {author} {\bibinfo {author} {\bibfnamefont {G.~J.}\ \bibnamefont
			{Doherty}}\ and\ \bibinfo {author} {\bibfnamefont {H.~T.}\ \bibnamefont
			{McMahon}},\ }\href {\doibase 10.1146/annurev.biochem.78.081307.110540}
	{\bibfield  {journal} {\bibinfo  {journal} {Annu. Rev. Biochem.}\ }\textbf
		{\bibinfo {volume} {78}},\ \bibinfo {pages} {857} (\bibinfo {year}
		{2009})}\BibitemShut {NoStop}%
	\bibitem [{\citenamefont {Ewers}\ and\ \citenamefont
		{Helenius}(2011)}]{Ewers:2011jj}%
	\BibitemOpen
	\bibfield  {author} {\bibinfo {author} {\bibfnamefont {H.}~\bibnamefont
			{Ewers}}\ and\ \bibinfo {author} {\bibfnamefont {A.}~\bibnamefont
			{Helenius}},\ }\href {\doibase 10.1101/cshperspect.a004721} {\bibfield
		{journal} {\bibinfo  {journal} {Cold Spring Harbor Perspectives in Biology}\
		}\textbf {\bibinfo {volume} {3}},\ \bibinfo {pages} {a004721} (\bibinfo
		{year} {2011})}\BibitemShut {NoStop}%
	\bibitem [{\citenamefont {Canton}\ and\ \citenamefont
		{Battaglia}(2012)}]{Canton:2012dl}%
	\BibitemOpen
	\bibfield  {author} {\bibinfo {author} {\bibfnamefont {I.}~\bibnamefont
			{Canton}}\ and\ \bibinfo {author} {\bibfnamefont {G.}~\bibnamefont
			{Battaglia}},\ }\href {\doibase 10.1039/c2cs15309b} {\bibfield  {journal}
		{\bibinfo  {journal} {Chem. Soc. Rev.}\ }\textbf {\bibinfo {volume} {41}},\
		\bibinfo {pages} {2718} (\bibinfo {year} {2012})}\BibitemShut {NoStop}%
	\bibitem [{\citenamefont {Kholodenko}(2006)}]{Kholodenko:2006cr}%
	\BibitemOpen
	\bibfield  {author} {\bibinfo {author} {\bibfnamefont {B.~N.}\ \bibnamefont
			{Kholodenko}},\ }\href {\doibase 10.1038/nrm1838} {\bibfield  {journal}
		{\bibinfo  {journal} {Nature}\ }\textbf {\bibinfo {volume} {7}},\ \bibinfo
		{pages} {165} (\bibinfo {year} {2006})}\BibitemShut {NoStop}%
	\bibitem [{\citenamefont {Sorkin}\ and\ \citenamefont {von
			Zastrow}(2009)}]{Sorkin:2009cb}%
	\BibitemOpen
	\bibfield  {author} {\bibinfo {author} {\bibfnamefont {A.}~\bibnamefont
			{Sorkin}}\ and\ \bibinfo {author} {\bibfnamefont {M.}~\bibnamefont {von
				Zastrow}},\ }\href {\doibase 10.1038/nrm2748} {\bibfield  {journal} {\bibinfo
			{journal} {Nat. Rev. Mol. Cell Biol.}\ }\textbf {\bibinfo {volume} {10}},\
		\bibinfo {pages} {609} (\bibinfo {year} {2009})}\BibitemShut {NoStop}%
	\bibitem [{\citenamefont {Sheetz}(2001)}]{Sheetz:2001ed}%
	\BibitemOpen
	\bibfield  {author} {\bibinfo {author} {\bibfnamefont {M.~P.}\ \bibnamefont
			{Sheetz}},\ }\href {\doibase 10.1038/35073095} {\bibfield  {journal}
		{\bibinfo  {journal} {Nat. Rev. Mol. Cell Biol.}\ }\textbf {\bibinfo {volume}
			{2}},\ \bibinfo {pages} {392} (\bibinfo {year} {2001})}\BibitemShut {NoStop}%
	\bibitem [{\citenamefont {Ananthakrishnan}\ and\ \citenamefont
		{Ehrlicher}(2007)}]{Ananthakrishnan:2007ux}%
	\BibitemOpen
	\bibfield  {author} {\bibinfo {author} {\bibfnamefont {R.}~\bibnamefont
			{Ananthakrishnan}}\ and\ \bibinfo {author} {\bibfnamefont {A.}~\bibnamefont
			{Ehrlicher}},\ }\href
	{http://eutils.ncbi.nlm.nih.gov/entrez/eutils/elink.fcgi?dbfrom=pubmed&id=17589565&retmode=ref&cmd=prlinks}
	{\bibfield  {journal} {\bibinfo  {journal} {Int. J. Biol. Sci.}\ }\textbf
		{\bibinfo {volume} {3}},\ \bibinfo {pages} {303} (\bibinfo {year}
		{2007})}\BibitemShut {NoStop}%
	\bibitem [{\citenamefont {Keren}(2011)}]{Keren:2011hg}%
	\BibitemOpen
	\bibfield  {author} {\bibinfo {author} {\bibfnamefont {K.}~\bibnamefont
			{Keren}},\ }\href {\doibase 10.1007/s00249-011-0741-0} {\bibfield  {journal}
		{\bibinfo  {journal} {Eur Biophys J}\ }\textbf {\bibinfo {volume} {40}},\
		\bibinfo {pages} {1013} (\bibinfo {year} {2011})}\BibitemShut {NoStop}%
	\bibitem [{\citenamefont {Chaikin}\ and\ \citenamefont
		{Lubensky}(2000)}]{Chaikin:2000td}%
	\BibitemOpen
	\bibfield  {author} {\bibinfo {author} {\bibfnamefont {P.~M.}\ \bibnamefont
			{Chaikin}}\ and\ \bibinfo {author} {\bibfnamefont {T.~C.}\ \bibnamefont
			{Lubensky}},\ }\href {http://books.google.co.in/books?id=P9YjNjzr9OIC} {\emph
		{\bibinfo {title} {{Principles of Condensed Matter Physics}}}}\ (\bibinfo
	{publisher} {Cambridge University Press},\ \bibinfo {year}
	{2000})\BibitemShut {NoStop}%
	\bibitem [{\citenamefont {Frenkel}\ and\ \citenamefont
		{Smit}(2001)}]{Frenkel:2001}%
	\BibitemOpen
	\bibfield  {author} {\bibinfo {author} {\bibfnamefont {D.}~\bibnamefont
			{Frenkel}}\ and\ \bibinfo {author} {\bibfnamefont {B.}~\bibnamefont {Smit}},\
	}\href {http://www.worldcat.org/isbn/0122673514} {\emph {\bibinfo {title}
		{{Understanding Molecular Simulation : From Algorithms to Applications}}}},\
\bibinfo {edition} {2nd}\ ed.\ (\bibinfo  {publisher} {Academic Press},\
\bibinfo {year} {2001})\BibitemShut {NoStop}%
\bibitem [{\citenamefont {Seifert}(1997)}]{Seifert:1997wq}%
\BibitemOpen
\bibfield  {author} {\bibinfo {author} {\bibfnamefont {U.}~\bibnamefont
		{Seifert}},\ }\href
{http://gateway.webofknowledge.com/gateway/Gateway.cgi?GWVersion=2&SrcAuth=mekentosj&SrcApp=Papers&DestLinkType=FullRecord&DestApp=WOS&KeyUT=A1997WE91800002}
{\bibfield  {journal} {\bibinfo  {journal} {Advances in Physics}\ }\textbf
	{\bibinfo {volume} {46}},\ \bibinfo {pages} {13} (\bibinfo {year}
	{1997})}\BibitemShut {NoStop}%
\bibitem [{\citenamefont {Tieleman}, \citenamefont {Marrink},\ and\
	\citenamefont {Berendsen}(1997)}]{Tieleman:1997ve}%
\BibitemOpen
\bibfield  {author} {\bibinfo {author} {\bibfnamefont {D.~P.}\ \bibnamefont
		{Tieleman}}, \bibinfo {author} {\bibfnamefont {S.-J.}\ \bibnamefont
		{Marrink}}, \ and\ \bibinfo {author} {\bibfnamefont {H.~J.}\ \bibnamefont
		{Berendsen}},\ }\href
{http://eutils.ncbi.nlm.nih.gov/entrez/eutils/elink.fcgi?dbfrom=pubmed&id=9512654&retmode=ref&cmd=prlinks}
{\bibfield  {journal} {\bibinfo  {journal} {Biochim Biophys Acta (BBA)
			-Reviews on Biomembranes}\ }\textbf {\bibinfo {volume} {1331}},\ \bibinfo
	{pages} {235} (\bibinfo {year} {1997})}\BibitemShut {NoStop}%
\bibitem [{\citenamefont {Venturoli}\ \emph {et~al.}(2006)\citenamefont
	{Venturoli}, \citenamefont {Maddalena~Sperotto}, \citenamefont {Kranenburg},\
	and\ \citenamefont {Smit}}]{Venturoli:2006kk}%
\BibitemOpen
\bibfield  {author} {\bibinfo {author} {\bibfnamefont {M.}~\bibnamefont
		{Venturoli}}, \bibinfo {author} {\bibfnamefont {M.}~\bibnamefont
		{Maddalena~Sperotto}}, \bibinfo {author} {\bibfnamefont {M.}~\bibnamefont
		{Kranenburg}}, \ and\ \bibinfo {author} {\bibfnamefont {B.}~\bibnamefont
		{Smit}},\ }\href {\doibase 10.1016/j.physrep.2006.07.006} {\bibfield
	{journal} {\bibinfo  {journal} {Physics Reports}\ }\textbf {\bibinfo {volume}
		{437}},\ \bibinfo {pages} {1} (\bibinfo {year} {2006})}\BibitemShut {NoStop}%
\bibitem [{\citenamefont {Ayton}\ and\ \citenamefont
	{Voth}(2010)}]{Ayton:2010gv}%
\BibitemOpen
\bibfield  {author} {\bibinfo {author} {\bibfnamefont {G.~S.}\ \bibnamefont
		{Ayton}}\ and\ \bibinfo {author} {\bibfnamefont {G.~A.}\ \bibnamefont
		{Voth}},\ }\href {\doibase 10.1016/j.semcdb.2009.11.011} {\bibfield
	{journal} {\bibinfo  {journal} {Seminars in Cell and Developmental Biology}\
	}\textbf {\bibinfo {volume} {21}},\ \bibinfo {pages} {357} (\bibinfo {year}
	{2010})}\BibitemShut {NoStop}%
\bibitem [{\citenamefont {Shinoda}, \citenamefont {DeVane},\ and\ \citenamefont
	{Klein}(2012)}]{Shinoda:2012ks}%
\BibitemOpen
\bibfield  {author} {\bibinfo {author} {\bibfnamefont {W.}~\bibnamefont
		{Shinoda}}, \bibinfo {author} {\bibfnamefont {R.}~\bibnamefont {DeVane}}, \
	and\ \bibinfo {author} {\bibfnamefont {M.~L.}\ \bibnamefont {Klein}},\ }\href
{\doibase 10.1016/j.sbi.2012.01.011} {\bibfield  {journal} {\bibinfo
		{journal} {Current Opinion in Structural Biology}\ }\textbf {\bibinfo
		{volume} {22}},\ \bibinfo {pages} {175} (\bibinfo {year} {2012})}\BibitemShut
{NoStop}%
\bibitem [{\citenamefont {Bradley}\ and\ \citenamefont
	{Radhakrishnan}(2013)}]{Bradley:2013gb}%
\BibitemOpen
\bibfield  {author} {\bibinfo {author} {\bibfnamefont {R.~P.}\ \bibnamefont
		{Bradley}}\ and\ \bibinfo {author} {\bibfnamefont {R.}~\bibnamefont
		{Radhakrishnan}},\ }\href {\doibase 10.3390/polym5030890} {\bibfield
	{journal} {\bibinfo  {journal} {Polymers}\ }\textbf {\bibinfo {volume} {5}},\
	\bibinfo {pages} {890} (\bibinfo {year} {2013})}\BibitemShut {NoStop}%
\bibitem [{\citenamefont {Ramakrishnan}, \citenamefont {Sunil~Kumar},\ and\
	\citenamefont {Radhakrishnan}(2014)}]{Ramakrishnan:2014ed}%
\BibitemOpen
\bibfield  {author} {\bibinfo {author} {\bibfnamefont {N.}~\bibnamefont
		{Ramakrishnan}}, \bibinfo {author} {\bibfnamefont {P.~B.}\ \bibnamefont
		{Sunil~Kumar}}, \ and\ \bibinfo {author} {\bibfnamefont {R.}~\bibnamefont
		{Radhakrishnan}},\ }\href {\doibase 10.1016/j.physrep.2014.05.001} {\bibfield
	{journal} {\bibinfo  {journal} {Physics Reports}\ }\textbf {\bibinfo
		{volume} {543}},\ \bibinfo {pages} {1} (\bibinfo {year} {2014})}\BibitemShut
{NoStop}%
\bibitem [{\citenamefont {Deserno}(2014)}]{Deserno:2014cv}%
\BibitemOpen
\bibfield  {author} {\bibinfo {author} {\bibfnamefont {M.}~\bibnamefont
		{Deserno}},\ }\href {\doibase 10.1016/j.chemphyslip.2014.05.001} {\bibfield
	{journal} {\bibinfo  {journal} {Chemistry and Physics of Lipids}\ } (\bibinfo
	{year} {2014}),\ 10.1016/j.chemphyslip.2014.05.001}\BibitemShut {NoStop}%
\bibitem [{\citenamefont {Canham}(1970)}]{Canham:1970p61}%
\BibitemOpen
\bibfield  {author} {\bibinfo {author} {\bibfnamefont {P.~B.}\ \bibnamefont
		{Canham}},\ }\href@noop {} {\bibfield  {journal} {\bibinfo  {journal} {J.
			Theor. Biol.}\ }\textbf {\bibinfo {volume} {26}},\ \bibinfo {pages} {61}
	(\bibinfo {year} {1970})}\BibitemShut {NoStop}%
\bibitem [{\citenamefont {Helfrich}(1973)}]{Helfrich:1973td}%
\BibitemOpen
\bibfield  {author} {\bibinfo {author} {\bibfnamefont {W.}~\bibnamefont
		{Helfrich}},\ }\href {http://www.ncbi.nlm.nih.gov/pubmed/4273690} {\bibfield
	{journal} {\bibinfo  {journal} {Z. Naturforsch. C}\ }\textbf {\bibinfo
		{volume} {28}},\ \bibinfo {pages} {693} (\bibinfo {year} {1973})}\BibitemShut
{NoStop}%
\bibitem [{\citenamefont {do~Carmo}(1976)}]{doCarmo:1976}%
\BibitemOpen
\bibfield  {author} {\bibinfo {author} {\bibfnamefont {M.~P.}\ \bibnamefont
		{do~Carmo}},\ }\href@noop {} {\emph {\bibinfo {title} {{Differential geometry
				of curves and surfaces}}}}\ (\bibinfo  {publisher} {Prentice Hall},\ \bibinfo
{address} {Engelwood Cliffs, New Jersey},\ \bibinfo {year}
{1976})\BibitemShut {NoStop}%
\bibitem [{\citenamefont {Shi}\ and\ \citenamefont
	{Baumgart}(2015)}]{Shi:2015ho}%
\BibitemOpen
\bibfield  {author} {\bibinfo {author} {\bibfnamefont {Z.}~\bibnamefont
		{Shi}}\ and\ \bibinfo {author} {\bibfnamefont {T.}~\bibnamefont {Baumgart}},\
}\href {\doibase 10.1038/ncomms6974} {\bibfield  {journal} {\bibinfo
	{journal} {Nat Comms}\ }\textbf {\bibinfo {volume} {6}},\ \bibinfo {pages}
{5974} (\bibinfo {year} {2015})}\BibitemShut {NoStop}%
\bibitem [{\citenamefont {Diz-Mu{\~n}oz}, \citenamefont {Fletcher},\ and\
	\citenamefont {Weiner}(2013)}]{DizMunoz:2013bi}%
\BibitemOpen
\bibfield  {author} {\bibinfo {author} {\bibfnamefont {A.}~\bibnamefont
		{Diz-Mu{\~n}oz}}, \bibinfo {author} {\bibfnamefont {D.~A.}\ \bibnamefont
		{Fletcher}}, \ and\ \bibinfo {author} {\bibfnamefont {O.~D.}\ \bibnamefont
		{Weiner}},\ }\href {\doibase 10.1016/j.tcb.2012.09.006} {\bibfield  {journal}
	{\bibinfo  {journal} {Trends in Cell Biology}\ }\textbf {\bibinfo {volume}
		{23}},\ \bibinfo {pages} {47} (\bibinfo {year} {2013})}\BibitemShut {NoStop}%

\bibitem [{\citenamefont {Nelson}\ and\ \citenamefont
	{Piran}(2004)}]{Nelson:2004vm}%
\BibitemOpen
\bibfield  {author} {\bibinfo {author} {\bibfnamefont {D.~R.}\ \bibnamefont
		{Nelson}}\ and\ \bibinfo {author} {\bibfnamefont {T.}~\bibnamefont {Piran}},\
}\href
{http://books.google.com/books?id=FbcMqgNrVjcC&pg=PA323&dq=intitle:Statistical+mechanics+of+membranes+and+surfaces&hl=&cd=1&source=gbs_api}
{{\emph {\bibinfo {title} {{Statistical Mechanics of Membranes and Surfaces}}}}}\ (\bibinfo  {publisher} {World Scientific},\
\bibinfo {year} {2004})\BibitemShut {NoStop}%

\bibitem [{\citenamefont {Ramakrishnan}, \citenamefont {Sunil~Kumar},\ and\
	\citenamefont {Ipsen}(2010)}]{Ramakrishnan:2010hk}%
\BibitemOpen
\bibfield  {author} {\bibinfo {author} {\bibfnamefont {N.}~\bibnamefont
		{Ramakrishnan}}, \bibinfo {author} {\bibfnamefont {P.~B.}\ \bibnamefont
		{Sunil~Kumar}}, \ and\ \bibinfo {author} {\bibfnamefont {J.~H.}\ \bibnamefont
		{Ipsen}},\ }\href {\doibase 10.1103/PhysRevE.81.041922} {\bibfield  {journal}
	{\bibinfo  {journal} {Phys. Rev. E}\ }\textbf {\bibinfo {volume} {81}},\
	\bibinfo {pages} {041922} (\bibinfo {year} {2010})}\BibitemShut {NoStop}%
\bibitem [{\citenamefont {Agrawal}, \citenamefont {Nukpezah},\ and\
	\citenamefont {Radhakrishnan}(2010)}]{Agrawal:2010iu}%
\BibitemOpen
\bibfield  {author} {\bibinfo {author} {\bibfnamefont {N.~J.}\ \bibnamefont
		{Agrawal}}, \bibinfo {author} {\bibfnamefont {J.}~\bibnamefont {Nukpezah}}, \
	and\ \bibinfo {author} {\bibfnamefont {R.}~\bibnamefont {Radhakrishnan}},\
}\href {\doibase 10.1371/journal.pcbi.1000926.s008} {\bibfield  {journal}
{\bibinfo  {journal} {PLoS Comput Biol}\ }\textbf {\bibinfo {volume} {6}},\
\bibinfo {pages} {e1000926} (\bibinfo {year} {2010})}\BibitemShut {NoStop}%
\bibitem [{\citenamefont {Ramanan}\ \emph {et~al.}(2011)\citenamefont
	{Ramanan}, \citenamefont {Agrawal}, \citenamefont {Liu}, \citenamefont
	{Engles}, \citenamefont {Toy},\ and\ \citenamefont
	{Radhakrishnan}}]{Ramanan:2011ds}%
\BibitemOpen
\bibfield  {author} {\bibinfo {author} {\bibfnamefont {V.}~\bibnamefont
		{Ramanan}}, \bibinfo {author} {\bibfnamefont {N.~J.}\ \bibnamefont
		{Agrawal}}, \bibinfo {author} {\bibfnamefont {J.}~\bibnamefont {Liu}},
	\bibinfo {author} {\bibfnamefont {S.}~\bibnamefont {Engles}}, \bibinfo
	{author} {\bibfnamefont {R.}~\bibnamefont {Toy}}, \ and\ \bibinfo {author}
	{\bibfnamefont {R.}~\bibnamefont {Radhakrishnan}},\ }\href {\doibase
	10.1039/c1ib00036e} {\bibfield  {journal} {\bibinfo  {journal} {Integr.
			Biol.}\ }\textbf {\bibinfo {volume} {3}},\ \bibinfo {pages} {803} (\bibinfo
	{year} {2011})}\BibitemShut {NoStop}%
\bibitem [{\citenamefont {Liu}\ \emph {et~al.}(2012)\citenamefont {Liu},
	\citenamefont {Tourdot}, \citenamefont {Ramanan}, \citenamefont {Agrawal},\
	and\ \citenamefont {Radhakrishanan}}]{Liu:2012es}%
\BibitemOpen
\bibfield  {author} {\bibinfo {author} {\bibfnamefont {J.}~\bibnamefont
		{Liu}}, \bibinfo {author} {\bibfnamefont {R.~W.}\ \bibnamefont {Tourdot}},
	\bibinfo {author} {\bibfnamefont {V.}~\bibnamefont {Ramanan}}, \bibinfo
	{author} {\bibfnamefont {N.~J.}\ \bibnamefont {Agrawal}}, \ and\ \bibinfo
	{author} {\bibfnamefont {R.}~\bibnamefont {Radhakrishanan}},\ }\href
{\doibase 10.1080/00268976.2012.664661} {\bibfield  {journal} {\bibinfo
		{journal} {Molecular Physics}\ }\textbf {\bibinfo {volume} {110}},\ \bibinfo
	{pages} {1127} (\bibinfo {year} {2012})}\BibitemShut {NoStop}%
\bibitem [{\citenamefont {Tourdot}, \citenamefont {Ramakrishnan},\ and\
	\citenamefont {Radhakrishnan}(2014)}]{Tourdot:2014wh}%
\BibitemOpen
\bibfield  {author} {\bibinfo {author} {\bibfnamefont {R.~W.}\ \bibnamefont
		{Tourdot}}, \bibinfo {author} {\bibfnamefont {N.}~\bibnamefont
		{Ramakrishnan}}, \ and\ \bibinfo {author} {\bibfnamefont {R.}~\bibnamefont
		{Radhakrishnan}},\ }\href
{http://journals.aps.org/pre/abstract/10.1103/PhysRevE.90.022717} {\bibfield
	{journal} {\bibinfo  {journal} {Phys. Rev. E}\ }\textbf {\bibinfo {volume}
		{90}},\ \bibinfo {pages} {022717} (\bibinfo {year} {2014})}\BibitemShut
{NoStop}%
\bibitem [{\citenamefont {Metropolis}\ \emph {et~al.}(1953)\citenamefont
	{Metropolis}, \citenamefont {Rosenbluth}, \citenamefont {Rosenbluth},
	\citenamefont {Teller},\ and\ \citenamefont {Teller}}]{Metropolis:1953in}%
\BibitemOpen
\bibfield  {author} {\bibinfo {author} {\bibfnamefont {N.}~\bibnamefont
		{Metropolis}}, \bibinfo {author} {\bibfnamefont {A.~W.}\ \bibnamefont
		{Rosenbluth}}, \bibinfo {author} {\bibfnamefont {M.~N.}\ \bibnamefont
		{Rosenbluth}}, \bibinfo {author} {\bibfnamefont {A.~H.}\ \bibnamefont
		{Teller}}, \ and\ \bibinfo {author} {\bibfnamefont {E.}~\bibnamefont
		{Teller}},\ }\href {\doibase 10.1063/1.1699114} {\bibfield  {journal}
	{\bibinfo  {journal} {J. Chem. Phys.}\ }\textbf {\bibinfo {volume} {21}},\
	\bibinfo {pages} {1087} (\bibinfo {year} {1953})}\BibitemShut {NoStop}%
\bibitem [{\citenamefont {Widom}(1963)}]{Widom:1963fl}%
\BibitemOpen
\bibfield  {author} {\bibinfo {author} {\bibfnamefont {B.}~\bibnamefont
		{Widom}},\ }\href {\doibase 10.1063/1.1734110} {\bibfield  {journal}
	{\bibinfo  {journal} {J. Chem. Phys.}\ }\textbf {\bibinfo {volume} {39}},\
	\bibinfo {pages} {2808} (\bibinfo {year} {1963})}\BibitemShut {NoStop}%
\bibitem [{\citenamefont {Bennett}(1976)}]{Bennett:1976gj}%
\BibitemOpen
\bibfield  {author} {\bibinfo {author} {\bibfnamefont {C.~H.}\ \bibnamefont
		{Bennett}},\ }\href {\doibase 10.1016/0021-9991(76)90078-4} {\bibfield
	{journal} {\bibinfo  {journal} {Journal of Computational Physics}\ }\textbf
	{\bibinfo {volume} {22}},\ \bibinfo {pages} {245} (\bibinfo {year}
	{1976})}\BibitemShut {NoStop}%
\bibitem [{\citenamefont {Roux}(1995)}]{Roux:1995vi}%
\BibitemOpen
\bibfield  {author} {\bibinfo {author} {\bibfnamefont {B.}~\bibnamefont
		{Roux}},\ }\href
{http://www.sciencedirect.com/science/article/pii/001046559500053I}
{\bibfield  {journal} {\bibinfo  {journal} {Computer Physics Communications}\
	}\textbf {\bibinfo {volume} {91}},\ \bibinfo {pages} {275} (\bibinfo {year}
	{1995})}\BibitemShut {NoStop}%
\bibitem [{\citenamefont {Tourdot}\ \emph {et~al.}(2014)\citenamefont
	{Tourdot}, \citenamefont {Bradley}, \citenamefont {Ramakrishnan},\ and\
	\citenamefont {Radhakrishnan}}]{Bradley:2014hm}%
\BibitemOpen
\bibfield  {author} {\bibinfo {author} {\bibfnamefont {R.~W.}\ \bibnamefont
		{Tourdot}}, \bibinfo {author} {\bibfnamefont {R.~P.}\ \bibnamefont
		{Bradley}}, \bibinfo {author} {\bibfnamefont {N.}~\bibnamefont
		{Ramakrishnan}}, \ and\ \bibinfo {author} {\bibfnamefont {R.}~\bibnamefont
		{Radhakrishnan}},\ }\href {\doibase 10.1049/iet-syb.2013.0057} {\bibfield
	{journal} {\bibinfo  {journal} {IET Systems Biology}\ }\textbf {\bibinfo
		{volume} {8}},\ \bibinfo {pages} {198} (\bibinfo {year} {2014})}\BibitemShut
{NoStop}%
\bibitem [{\citenamefont {Ramakrishnan}\ \emph {et~al.}()\citenamefont
	{Ramakrishnan}, \citenamefont {Eckmann}, \citenamefont {Ayyaswamy},
	\citenamefont {Weaver},\ and\ \citenamefont
	{Radhakrishnan}}]{Ramakrishnan:ureview}%
\BibitemOpen
\bibfield  {author} {\bibinfo {author} {\bibfnamefont {N.}~\bibnamefont
		{Ramakrishnan}}, \bibinfo {author} {\bibfnamefont {D.~M.}\ \bibnamefont
		{Eckmann}}, \bibinfo {author} {\bibfnamefont {P.~S.}\ \bibnamefont
		{Ayyaswamy}}, \bibinfo {author} {\bibfnamefont {V.~M.}\ \bibnamefont
		{Weaver}}, \ and\ \bibinfo {author} {\bibfnamefont {R.}~\bibnamefont
		{Radhakrishnan}},\ }\href@noop {} {\bibinfo  {journal} {Under Review}\
}\BibitemShut {NoStop}%
\bibitem [{\citenamefont {Souaille}\ and\ \citenamefont
	{Roux}(2001)}]{Souaille:2001gm}%
\BibitemOpen
\bibfield  {journal} {  }\bibfield  {author} {\bibinfo {author} {\bibfnamefont
		{M.}~\bibnamefont {Souaille}}\ and\ \bibinfo {author} {\bibfnamefont
		{B.}~\bibnamefont {Roux}},\ }\href {\doibase 10.1016/S0010-4655(00)00215-0}
{\bibfield  {journal} {\bibinfo  {journal} {Computer Physics Communications}\
	}\textbf {\bibinfo {volume} {135}},\ \bibinfo {pages} {40} (\bibinfo {year}
	{2001})}\BibitemShut {NoStop}%
\bibitem [{\citenamefont {Liu}\ \emph {et~al.}(2010)\citenamefont {Liu},
	\citenamefont {Weller}, \citenamefont {Zern}, \citenamefont {Ayyaswamy},
	\citenamefont {Eckmann}, \citenamefont {Muzykantov},\ and\ \citenamefont
	{Radhakrishnan}}]{Liu:2010em}%
\BibitemOpen
\bibfield  {author} {\bibinfo {author} {\bibfnamefont {J.}~\bibnamefont
		{Liu}}, \bibinfo {author} {\bibfnamefont {G.~E.}\ \bibnamefont {Weller}},
	\bibinfo {author} {\bibfnamefont {B.}~\bibnamefont {Zern}}, \bibinfo {author}
	{\bibfnamefont {P.~S.}\ \bibnamefont {Ayyaswamy}}, \bibinfo {author}
	{\bibfnamefont {D.~M.}\ \bibnamefont {Eckmann}}, \bibinfo {author}
	{\bibfnamefont {V.~R.}\ \bibnamefont {Muzykantov}}, \ and\ \bibinfo {author}
	{\bibfnamefont {R.}~\bibnamefont {Radhakrishnan}},\ }\href {\doibase
	10.1073/pnas.1006611107/-/DCSupplemental} {\bibfield  {journal} {\bibinfo
		{journal} {Proc. Natl. Acad. Sci. U.S.A.}\ }\textbf {\bibinfo {volume}
		{107}},\ \bibinfo {pages} {16530} (\bibinfo {year} {2010})}\BibitemShut
{NoStop}%
\bibitem [{\citenamefont {Liu}\ \emph {et~al.}(2011)\citenamefont {Liu},
	\citenamefont {Agrawal}, \citenamefont {Calderon}, \citenamefont {Ayyaswamy},
	\citenamefont {Eckmann},\ and\ \citenamefont {Radhakrishnan}}]{Liu:2011fe}%
\BibitemOpen
\bibfield  {author} {\bibinfo {author} {\bibfnamefont {J.}~\bibnamefont
		{Liu}}, \bibinfo {author} {\bibfnamefont {N.~J.}\ \bibnamefont {Agrawal}},
	\bibinfo {author} {\bibfnamefont {A.}~\bibnamefont {Calderon}}, \bibinfo
	{author} {\bibfnamefont {P.~S.}\ \bibnamefont {Ayyaswamy}}, \bibinfo {author}
	{\bibfnamefont {D.~M.}\ \bibnamefont {Eckmann}}, \ and\ \bibinfo {author}
	{\bibfnamefont {R.}~\bibnamefont {Radhakrishnan}},\ }\href {\doibase
	10.1016/j.bpj.2011.05.063} {\bibfield  {journal} {\bibinfo  {journal}
		{Biophys. J.}\ }\textbf {\bibinfo {volume} {101}},\ \bibinfo {pages} {319}
	(\bibinfo {year} {2011})}\BibitemShut {NoStop}%
\bibitem [{Note1()}]{Note1}%
\BibitemOpen
\bibinfo {note} {Calculated as (nanocarrier radius + length of the antibody +
	length of the receptor + $d_{0}$)}\BibitemShut {NoStop}%
\end{thebibliography}
%

\end{document}